\documentstyle{article}
\newfont{\g}{eufm9}

\newcommand{\gtg}{\mbox{\g g}}

\newcommand{\gtb}{\mbox{\g b}}
\newcommand{\hgtb}{\mbox{$\hat{\gtb}$}}
\newcommand{\hgtg}{\mbox{$\hat{\gtg}$}}
\newcommand{\hgth}{\mbox{$\hat{\gth}$}}

\newcommand{\gtsl}{\mbox{\g sl}}
\newcommand{\gtgl}{\mbox{\g gl}}
\newcommand{\gtn}{\mbox{\g n}}
\newcommand{\gtnp}{\gtn_{+}}
\newcommand{\gtnm}{\gtn_{-}}
\newcommand{\gth}{\mbox{\g h}}
\newcommand{\nc}{\mbox{${\bf C}$}}
\newcommand{\dq}{\mbox{$\partial_{q}$}}
\newcommand{\tdq}{\mbox{$\tilde{\partial_{q}}$}}
\newcommand{\nz} {\mbox{${\bf Z}$}}
\newcommand{\nn} {\mbox{${\bf N}$}}
\newcommand{\cp} {\mbox{${\bf CP}$}}
\newcommand{\crr}{\mbox{${\cal R}$}}
\newcommand{\tg}{\mbox{$\tilde {\Gamma}$}}

\newcommand{\co}{\mbox{${\cal O}$}}

\newcommand{\ch}{\mbox{${\cal H}$}}
\newcommand{\cl}{\mbox{${\cal L}$}}

\newcommand{\cm}{\mbox{${\cal M}$}}
\newcommand{\cv}{\mbox{${\cal V}$}}
\newcommand{\ca}{\mbox{${\cal A}$}}

\newcommand{\binq}{\mbox{$\left[ \begin{array}{c}n\\j \end{array}\right]_{d}$}}

\newtheorem{theorem}{Theorem}[section]
\newtheorem{example}[theorem]{Example}

\newtheorem{lemma}[theorem]{Lemma}\title{{ \bf Integral  Intertwining Operators
and Complex Powers of Differential ($q-$Difference) Operators}}
\author{Boris Feigin\\
Landau Institute for Theoretical Physics
\and
Feodor Malikov\thanks{Supported by the
Japan Society for the Promotion of Science Post -Doctoral Fellowship for
Foreign Researchers in
Japan.}\\
 Department of Mathematics, Kyoto University ,\\
 Kyoto 606 Japan }
\date{Received: }
 
\begin{document}

\maketitle

\begin{abstract}
We study a family of modules over Kac-Moody algebras realized in multi-valued
functions on
a flag manifold and find integral representations for intertwining operators
acting on these
modules. These intertwiners are related to some expressions involving complex
powers of  Lie algebra
generators.
 When applied to affine Lie algebras, these expressions give integral formulas
for correlation
functions with values in not necessarily highest weight modules.
 We write related formulas out
in an explicit form in the case of $\hat{\gtsl_{2}}$. The latter formulas admit
q-deformation
producing an integral representation of q-correlation functions. We also
discuss a relation
of complex
powers of  Lie algebra (quantum group) generators and Casimir operators to
($q-$)special functions.

\vspace{5 mm}

\end{abstract}
\section{{\bf Introduction}}

This paper consists of 2 parts independent within reasonable limits. The 1st
one
(sect. \ref{Construction_of_Intertwining_Operators}) is devoted to constructing
 integral
intertwining operators acting between  Kac-Moody Lie algebra modules realized
in multi-valued
functions on a flag manifold. The 2nd (sect.
\ref{Solutions_of_Knizhnik_Zamolodchikov_Equations_The_classical_case},
\ref{Solutions_of_q_Knizhnik_Zamolodchikov_Equations}) is devoted to integral
representations of
solutions to
the (quantum) Knizhnik-Zamolodchikov equation. Both have in common, firstly,
Kac-Moody Lie algebra
(quantum group) singular vector formula as a motivation and, secondly, a new
class of Kac-Moody Lie
algebra modules as a main object.

\subsection{{\bf Intertwining Operators}}

1. Let $\gtg$ be a Kac-Moody Lie agebra on canonical generators
$F_{i},H_{i}, E_{i},\;0\leq i\leq n$ and defining relations, associated to a
generalized
symmetrizable Cartan
matrix $A=(a_{ij})$:
\[ [h_{i},h_{j}]=0,\;[h_{i},E_{j}]=a_{ij}E_{j},\;[h_{i},F_{j}]=-a_{ij}F_{j},\]
\[ad^{-a_{ij}+1}(E_{i})E_{j}=ad^{-a_{ij}+1}(F_{i})F_{j}=0.\]
These relations admit the antiinvolution $\omega$ such that: $\omega (E_{i})=
F_{i},\;\omega (F_{i})
=E_{i},\;\omega (H_{i})=H_{i}$.
Denote by  $\gtnp$ ($\gtnm$ resp.) the subalgebra generated by
$E_{0},\ldots,E_{n}$
($F_{0},\ldots,F_{n}$ resp.) and by $\gth$ - the one generated by
$H_{0},\ldots,H_{n}$. Denote also
by $M(\lambda)$ the Verma module  over $\gtg$ with the highest weight
$\lambda\in \gth^{\ast}$.
This is a module on 1 generator
$v_{\lambda}$ and defining relations:
\[ \gtnp v_{\lambda}=0,\;H_{i}v_{\lambda}=\lambda (H_{i}).\]
We will also be using a contragredient Verma module $M(\lambda)^c$ which is
defined as follows. As a
vector space $M(\lambda)^c$ is dual to $M(\lambda)$ but the action of $\gtg$ is
different from the
canonical action on a dual space: if $f(.)\in M(\lambda)^{c}$ is a linear
functional on $M(\lambda)$
and $g\in\gtg$ then we set $gf(.)=f(\omega (g).)$. Obviously a map dual to a
morphism of Verma
modules is a morphism of contragredient Verma modules.

A $\gtg-$ morphism of $M(\lambda)$ is uniquely determined by the image of
$v_{\lambda}$. It follows
from the definition that a
non-zero vector $w$ of a $\gtg-$module $W$ may serve as an image of
$v_{\lambda}$
under a non-zero morphism $M(\lambda)\rightarrow W$ if and only if it satisfies
the same conditions
as $v_{\lambda}$:
\begin{equation}
\gtnp v_{\lambda}=0,\;H_{i}v_{\lambda}=\lambda (H_{i}).
\label{def_sing_vect}
\end{equation}
A non-zero element $w$ of a $\gtg-$ module $W$ is said to be a singular vector
of the
 weight $\lambda$
if it satisfies (\ref{def_sing_vect}). It follows that the problem of
classification of morphisms of
Verma modules into a given  module is equivalent to that of singular vectors in
the latter module.
We now recall a singular vector formula obtained in \cite{malff}.

Let $(.,.)$ be an invariant inner product on $\gth^{\ast}$, $\Delta$ be the set
of roots of
$\gtg$, $\alpha_{0},\ldots,\alpha_{n}$ be the set of simple roots related to
the generators
$E_{0},\ldots,E_{n}$. A root is called positive if it is equal to a
non-negative
linear combination
of simple roots.
To each simple root we associate a reflection of the space $\gth^{\ast}$
\[r_{i}\lambda=\lambda - \frac{2(\lambda
,\alpha_{i})}{(\alpha_{i},\alpha_{i})}\alpha_{i}.\]
The group generated by these reflections is called the Weyl group of $\gtg$.
Its action preserves
the
set of roots and one defines the set of real roots as
$\Delta^{re}=W\{\alpha_{0},\ldots,\alpha_{n}\}$.

 We define the shifted action of the Weyl group $W$ on $\gth^{\ast}$ by
\[r_{\alpha}\cdot\lambda=\lambda - \frac{2(\lambda +\rho
,\alpha)}{(\alpha,\alpha)}\alpha,\]
where $r_{\alpha}$ stands for the reflection at a real root $
\alpha$ and $\rho$ is the fixed element of $\gth^{\ast}$ determined by the
following conditions
$\rho (H_{0})=\ldots =\rho (H_{n})=1$.

 The Kac-Kazhdan determinant formula \cite{kac_kazhd} implies that for a
positive
real root $\alpha$ the Verma
module $M(\lambda)$ contains a singular vector of the weight
$r_{\alpha}\cdot\lambda$ provided
\begin{equation}
\frac{2(\lambda +\rho ,\alpha)}{(\alpha,\alpha)}\subset\{1,2,\ldots\}.
\label{kac_k_cond}
\end{equation}
It also follows from \cite{kac_kazhd}
that for a generic $\lambda$ satisfying  this condition the singular vector is
unique up to
proportionality. The formula for this singular vector was found in \cite{malff}
 in the following form.
\begin{theorem}
\label{m_f_f}

Let
$r_{\alpha}=r_{i_{l}}\cdots r_{i_{2}}r_{i_{1}}$ be a reduced decomposition. Set

\[\beta_{j}=\frac{2(r_{i_{j-1}}\cdots r_{i_{1}}\cdot\lambda,\alpha_{i_{j}})}
{(\alpha_{i_{j}},\alpha_{i_{j}})}+1.\]

Then provided (\ref{kac_k_cond}) holds the vector
\begin {equation}
F_{i_{l}}^{\beta_{l}}\cdots
F_{i_{2}}^{\beta_{2}}F_{i_{1}}^{\beta_{1}}v_{\lambda}
\label{sing_vect_form}
\end{equation}
is singular.
\end{theorem}
This singular vector formula involves complex powers of
Lie algebra generators and therefore, its meaning has still to
be clarified. Here we restrict ourselves to the
following comment. Regard the expression $F_{i_{l}}^{j_{l}}\cdots
F_{i_{2}}^{j_{2}}F_{i_{1}}^{j_{1}}$
as a function of integers $(j_{1},\ldots,j_{k})\in\nz^{k}$
taking values in the universal enveloping algebra
$U(\gtnm)$. One shows that it admits an analytic continuation to a function
defined on $\nc^{k}$
taking values in a certain completion of $U(\gtnm)$. Then one shows that its
value on a sequence
$(\beta_{1},\ldots,\beta_{l})$ defined in Theorem \ref{m_f_f} actually belongs
to $U(\gtnm)$,
provided (\ref{kac_k_cond}) holds.

This explanation obscures the fact that the intertwiner related to the singular
vector
(\ref{sing_vect_form}) is a composition of intertwiners which  exist even if
 (\ref{kac_k_cond}) does not hold, though may not act between Verma
modules. These intertwiners can be easily constructed once one has defined an
action of complex
powers of the Lie algebra generators in a way that some natural conditions are
satisfied.
 Consider, for example, the expression $F_{i_{1}}^{\beta_{1}}v_{\lambda}$. If
one has convinced
himself that the easily checked relation
\begin{equation}
[E_{i}
,F_{j}^{\beta}]=\delta_{i,j}\,\beta\,F_{j}^{\beta-1}(H_{i}-\beta+1),\;\beta\in\nn
\label{comm_s_v_0}
\end{equation}
makes sense for $\beta\in\nc$, one obtains
\begin{equation}
E_{j}F_{i_{1}}^{\beta_{1}}v_{\lambda}=
\delta_{j,i_{1}}\beta_{1}F_{i_{1}}^{\beta_{1}-1}(H_{i}-\beta_{1}+1)v_{\lambda}=0.
\label{comm_s_v}
\end{equation}
Therefore, the vector

\begin{equation}
F_{i_{1}}^{\beta_{1}}v_{\lambda}=F_{i_{1}}^{\lambda(H_{i_{1}})+1}v_{\lambda}
\label{simplestsingvectorintr}
\end{equation}

 is a singular vector of a module which we
have not yet constructed. The same reasoning  for $v_{\lambda}$ replaced with
 $F_{i_{1}}^{\beta_{1}}v_{\lambda}$
shows that the vector $F_{i_{2}}^{\beta_{2}}F_{i_{1}}^{\beta_{1}}v_{\lambda}$
is singular. This
procedure can be iterated arbitrary number of times, always resulting in  a
singular vector.
So provided with a
suitable definition of a complex power of a Lie algebra generator one can
associate with a Weyl
group element $w$ and its  decomposition
$w=r_{i_{l}}\cdots r_{i_{2}}r_{i_{1}}$ a singular vector
(\ref{sing_vect_form}).

The purpose of this work was to provide this construction with
 a rigid foundation. We define a family of
$\gtg-$modules, realized in functions on the big cell of the corresponding flag
manifold
 so that the complex powers of Cartan generators of $\gtg$ are well-defined
integral
operators,
acting usually from one module into another.

2.  It is known that a contragredient  Verma
module  is realized in sections of a line bundle over the  flag manifold
with singularities at Schubert cells of codimension $\geq 1$,  $\gtg$ acting by
vector fields. Thus
in order to define a complex power of a Cartan generator one has first
to define a complex power of a
vector field as an operator acting on functions.
 Linearizing a vector field in a neighborhood of a non-singular point, one
reduces the
problem to evaluation of a quantity $(\frac{d}{dz})^{\mu}f(z)$ for an analytic
(probably
multi-valued) function $f(z)$ of a complex variable $z$ and an arbitrary
$\mu\in\nc$. But this is a
classical topic. If $\mu\in\nn$ then the answer is given by the Cauchy
integral
\[(\frac{d}{dz})^{\mu}f(z)=\frac{1}{\mu !}\oint_{\sigma}\frac{f(t)}{(t-z)^{\mu
+1}}dt,\]
where $f(t)$ is supposed to be regular at $t=z$ and $\sigma$ is a sufficiently
small contour
enclosing the point $t=z$. In order to get an analytic continuation of a Cauchy
integral (over $\mu$)
one has to demand that the function $f(t)$ is branching at some point, say at
$t=0$. Then up to
some factor (depending only on a branching coefficient of $f(t)$ at $t=0$) one
has
\begin {equation}
\label{compl_der}
(\frac{d}{dz})^{\mu}f(z)=\frac{1}{\Gamma (\mu+1) }
\oint_{\sigma_{1}\sigma_{2}\sigma_{1}^{-1}\sigma_{2}^{-1}}
\frac{f(t)}{(t-z)^{\mu +1}}dt,
\end{equation}
where $\Gamma (.)$ is the $\Gamma-$function and $\sigma_{1},\sigma_{2}$ are
generators of the 1st
homotopy group of $\nc - \{0,z\}$.
The formula (\ref{compl_der}) is sufficient to describe our construction in the
case of $\gtsl_{2}$
which is also one of the cornerstones of the general case.

3.{\bf The $\gtsl_{2}-$case.}
The flag manifold in this case is $\cp^{1}$, the big cell is $\nc$. The Cartan
generators are
realized as follows:

\begin{equation}
\label{realsl2}
 E=-\frac{d}{dz},\; H=-2z\frac{d}{dz},\; F=z^{2}\frac{d}{dz},
\end{equation}

 where $z$ is a
coordinate on $\nc$. The algebra of  all vector fields acts  on tensor fields
of the form
$f(z)dz^{-\lambda /2},\;f(z)\in\nc[z,z^{-1}]z^{\nu},\nu\in\nc,$ as follows

\begin{equation}
\label{act_tens}
p(z)\frac{d}{dz}(f(z)dz^{-\lambda /2}=(p(z)f'(z)-\lambda /2
p'(z)f(z))dz^{-\lambda /2}.
\end{equation}

Specifying the class of functions $f(z)$ one obtains the 2-parametric family of
$\gtsl_{2}-$modules
given by

\begin{equation}
\label{fam_sl2_mod}
V(\nu ,\lambda)=z^{\nu} \nc [z,z^{-1}]dz^{-\lambda/2}.
\end{equation}

It is worth mentioning that $V(0,\lambda
)$  contains the
submodule $\nc [z]dz^{-\lambda /2}$ isomorphic to a contragredient Verma module
with
the highest weight $\lambda$. The latter is irreducible if $\lambda$ is not a
nonnegative integer and
contains the $\lambda +1$-dimensional irreducible $\gtsl_{2}-$submodule spanned
by
$\{dz^{-\lambda /2},zdz^{-\lambda /2},\ldots , z^{\lambda }dz^{-\lambda /2}\}$
otherwise.

 Now consider a meromorphic tensor field over $\cp^{1}\times\cp^{1}$:
\[s=\frac{dt\,dx}{(t-x)^{2}}\;.\]
A direct calculation shows that $s$ is $\gtg-$invariant. Then take a power
$s^{\lambda/2+1}$.
Whatever the meaning of this expression is for a complex $\lambda$, the result
is also
$\gtg-$invariant, and one considers the product
\[ f(t)\,dt^{-\lambda/2}\,s^{\lambda/2+1}=\{\frac{f(t)}{(t-z)^{\lambda
+2}}dt\}dz^{\lambda /2+1}.\]
The result admits integration as in (\ref{compl_der}), which defines the
operator

\begin{eqnarray}
T;\;V(\nu ,\lambda)&\rightarrow&
V(\nu-\lambda,-\lambda-2)\label{int_sl2_form}\\
T:\;f(x)\,dx^{-\lambda/2}&\mapsto&\{\frac{1}{\Gamma (\lambda +2) }
\oint_{\sigma_{1}\sigma_{2}\sigma_{1}^{-1}\sigma_{2}^{-1}}
\frac{f(t)}{(t-x)^{\lambda +2}}dt\}dx^{\lambda /2+1}\nonumber
\end{eqnarray}

We have arrived at

\begin{lemma} The operator $T$ as in (\ref{int_sl2_form}
) is $\gtsl_{2}-$linear.
\label{int_sl2}
\end{lemma}

Note that if $\lambda$ is a nonnegative integer then
the restriction of our intertwiner to the
submodule $\nc [z]dz^{-\lambda /2}\subset V(0,\lambda)$ is a map dual to the
one determined by the singular vector
$F^{\lambda+1}v_{\lambda}$ of the Theorem\ref{m_f_f} and it acts killing the
irreducible submodule
spanned by
$\{dz^{-\lambda /2},zdz^{-\lambda /2},\ldots , z^{\lambda }dz^{-\lambda /2}\}$
and establishing
the isomorphism of the correponding quotient with the (irreducible) module
\mbox{$V(0,-\lambda-2)$}.

4. The transparent construction of Lemma\ref{int_sl2} admits a generalization
to the case of a simple
finite-dimensional Lie algebra or non-twisted affine Lie algebra. One
associates  in a way
compatible with the action of a group $G$ ( related to $\gtg$ )
to each point $x$ on a flag manifold and a fixed
simple root $\alpha$ - a projective
line $X_{\alpha}(x)$ ( closed Schubert cell related to this root and attached
to this point).
Then one consructs $s_{i}$ ( an analogue of
$s$ of the $\gtsl_{2}$-case)
: a tensor field on the cartesian product of 2 copies of a
 flag manifold having good properties with respect to the action of $G$
and such that its restriction to each 1-dimensional Schubert
 cell attached  to each point is a section of the line
bundle of degree -2. Then one takes an element of an appropriate $\gtg-$module;
realizes it,
roughly speaking, as a multivalued section of a bundle induced from a character
$\lambda$
 of the maximal torus
; multiplies it by an appropriate power of $s_{i}$ so that the restriction of
the result to each
 1-dimensinal Schubert cell is a differential form and, finally, integrates it
over each
1-dimensional Schubert cell. What one gets is a multi-valued section of a line
bundle induced
from $r_{\alpha}\cdot\lambda$.

This operator is closely related to the singular vector associated to the
simple reflection. One
may now associate to an arbitrary element $w$ of the
Weyl group( or rather to its reduced decomposition)
the corresponding composition of constructed intertwiners. But there are some
other possibilities. As
suggested by the formula (\ref{int_sl2_form}) an intertwiner related to an
element $w$
of the Weyl group
is an integration over a cycle of the Schubert cell $X_{w}(x)$ with
coefficients in a certain local
system. Taking composition means a special choice of the cycle. There should be
other cycles as well,
and there is some evidence suggested by the developments in conformal field
theory
and geometry ( see, for
example, ~\cite{bowkn} , ~\cite{sch_varch_2}
 and also see below) that these cycles should be related to a certain quantum
group.

\subsection {{\bf Integral Formulas for Correlation Functions}}
 Let $\gtg$ be a finite-dimensional simple Lie algebra, $\hgtg$ the
corresponding non-twisted affine
Lie algebra. Let $\lambda_{1},\lambda_{2}$ be weights of $\gtg$ ,
$M(\lambda_{1},\,  k)$ ($M(\lambda_{2},\, k)^{c}$ ) be  the Verma (
contragredient Verma) module
over $\hgtg$ with the central charge $k$;
let also $V$ be a $\gtg-$module and $V((z))$ the module of formal Laurent
series in $z$ with coefficients
in $V$, regarded as a $\hgtg-$module with the central charge equal to 0.

Vertex operator is a $\hgtg-$linear map
\[\Phi (z):\; M
(\lambda_{1},\,  k)\rightarrow
 M(\lambda_{2},\, k)^{c}\otimes V((z)).\]

One may consider a product of vertex operators
$\Phi_{m}(z_{m})\circ\cdots\circ\Phi_{1}(z_{1})$
and matrix
elements $\langle w^{\ast},\, \Phi_{m}(z_{m})\circ\cdots\circ\Phi_{1}(z_{1})
v\rangle$, where
$v\in M(\lambda_{1},\,  k),\; w^{\ast}\in (M(\lambda_{m+1},\, k)^{c})^{\ast}$.
One of the central
results of conformal field theory is ( see ~\cite{kn_zam})

\begin{theorem}[Knizhnik, Zamolodchikov]
The matrix element related to a pair of vacuum vectors
\[\Psi (z)= \langle v_{\lambda_{m+1}}^{\ast}\,
\Phi_{m}(z_{m})\circ\cdots\circ\Phi_{1}(z_{1})
v_{\lambda_{1}}\rangle\]
satisfies a certain system of
differential equations.
\end{theorem}

The explicit form of
the above-mentioned system of differential equations called
system of Knizhnik-Zamolodchikov equations may be found in
sect.\ref{tr_f_k_z_eq}.

One shows that given
\[\Psi (z)= \langle v_{\lambda_{m+1}}^{\ast}\,
\Phi_{m}(z_{m})\circ\cdots\circ\Phi_{1}(z_{1})
v_{\lambda_{1}}\rangle\]
the matrix element
\[ \langle v_{\lambda_{m}}^{\ast}\,
\Phi_{m}(z_{m})\circ\cdots\circ\Phi_{1}(z_{1})
F_{i_{l}}^{\beta_{l}}\cdots F_{i_{2}}^{\beta_{2}}F_{i_{1}}^{\beta_{1}}
v_{\lambda_{1}}
\rangle\]
is equal to

\[F_{i_{1}}^{\beta_{1}}F_{i_{2}}^{\beta_{2}}\cdots F_{i_{l}}^{\beta_{l}}\,\Psi
(z).\]
Suppose  we are able to define an action of complex powers of $F_{i}'$s (it
should seem plausible
now). Then if one chooses exponents $\beta_{j}$ as prescribed by the singular
vector formula
 (\ref{sing_vect_form}), the vector
$F_{i_{l}}^{s_{l}}\cdots F_{i_{2}}^{s_{2}}F_{i_{1}}^{s_{1}} v_{\lambda_{1}}$ is
(at least
formally)  singular and one may hope that the expression
\[F_{i_{1}}^{\beta_{1}}F_{i_{2}}^{\beta_{2}}\cdots F_{i_{l}}^{\beta_{l}}\,\Psi
(z)\]
also satisfies the Knizhnik-Zamolodchikov equations. We prove that it is true
for the usual
definition

\begin{eqnarray}
F_{i_{1}}^{\beta_{1}}F_{i_{2}}^{\beta_{2}}\cdots F_{i_{l}}^{\beta_{l}}\,\Psi
(z) =
&\prod_{j=1}^{l}& \Gamma
 (-\beta_{j})^{-1}\times \nonumber\\
 \int \{\exp (-t_{1}F_{i_{1}})\cdots\exp (-t_{l}F_{i_{l}}) \Psi\}
&\prod_{j=1}^{l}&t_{j}
^{-\beta_{j}-1}\,dt_{1}\cdots dt_{l},\label{compl_pow_def}
\end{eqnarray}

where the cycle of integration is defined to be any element of the highest
homology group with coefficients in the
local system of single-valued branches of the integrand.

\subsection {{\bf The Quantum Case}}
The Jimbo - Drinfeld quantum group  $U_{q}(\gtg)$ is a Hopf algebra on
generators
\mbox{$E_{i},\;F_{i},\;K_{i}^{\pm 1},\;1\leq i\leq n$}
and relations, which represent a deformation of those
for the Kac-Moody Lie algebra $\gtg$, $q$ being a parameter of deformation.
Considerable parts
of the
representation theory of $U_{q}(\gtg)$ and $\gtg$ are parallel to each other.
 This, in particular,
enabled I.Frenkel and N.Reshetikhin ( see ~\cite{fr_resh} )
to give definitions of a $q-$vertex operator, correlation
function and derive a system of $q-$difference equations ($q$KZ equations)
satisfied by  the
latter.

The same can be said about the representation theory of highest weight modules.
(See, for example, ~\cite{dc_kac} ,~\cite{lus} )
Everything discussed
above in the $\gtg-$case
admits a direct $q-$deformation.
For example, the following simple relation (c.f. formula (\ref{comm_s_v_0}) )
\[[E_{i},F_{j}^{s}]=\delta_{i,j}
\frac{q_{j}^{s}-q_{j}^{-s}}{q_{j}-q_{j}^{-1}}\,F_{j}^{s-1}
\frac{K_{j}q_{j}^{-s+1}-K_{j}^{-1}q_{j}^{s-1}}{q_{j}-q_{j}^{-1}}\]
together with a calculation analogous to that in (\ref {comm_s_v})
implies that the formula, literally coinciding with (\ref{sing_vect_form})
gives, at least formally,
a singular vector in a  module over $U_{q}(\gtg)$. (See \cite{mal} for a
precise statement analogous
to Theorem \ref{m_f_f}.) This makes it plausible that the formula of the type
\[F_{i_{1}}^{\beta_{1}}F_{i_{2}}^{\beta_{2}}\cdots F_{i_{l}}^{\beta_{l}}\,\Psi
(z)\]
gives a solution of $q$KZ equations for a given solution $\Psi (z)$. We prove
that it is true provided
\[F_{i_{1}}^{\beta_{1}}F_{i_{2}}^{\beta_{2}}\cdots F_{i_{l}}^{\beta_{l}}\,\Psi
(z)\] is defined by
\begin{eqnarray}
F_{i_{1}}^{\beta_{1}}F_{i_{2}}^{\beta_{2}}\cdots F_{i_{l}}^{\beta_{l}}\,\Psi
(z) =
&\prod_{j=1}^{l}& \Gamma_{q}
 (-\beta_{j})^{-1}\times \nonumber\\
 \int \{\exp_{q} (-t_{1}F_{i_{1}})\cdots\exp_{q} (-t_{l}F_{i_{l}}) \Psi\}
&\prod_{j=1}^{l}&t_{j}
^{-\beta_{j}-1}\,d_{q}t_{1}\cdots d_{q}t_{l},\label{compl_pow_def_quant}
\end{eqnarray}

Note that (\ref{compl_pow_def_quant}) is completely analogous to
(\ref{compl_pow_def}) with all the
ingredients replaced with their $q-$analogs: Gamma function with $q-$Gamma
function, exponential
function with $q-$exponential function, and integral with $q-$integral (
Jackson integral ).

There is also a problem of convergence of the integral in
(\ref{compl_pow_def_quant}). The classical
counterpart of this problem is solved by considering the realization of
$\gtg-$modules in analytic
functions on a flag manifold. We do the same for $U_{q}(\gtsl_{2})$
constructing a 2-parametric
family of actions of $U_{q}(\gtsl_{2})$ by difference operators, which deforms
the family of
$\gtsl_{2}-$modules $V(\mu,\lambda)$. This also leads to explicit formulas for
solutions of
$q$KZ equations for $U_{q}(\widehat{\gtsl_{2}})$.

{\bf Remarks.} (i) Note that usually KZ equations are considered for
finite-dimensional, or at
least, highest weight $\gtg-$modules $V_{i}'$s. This is not the case here.
Therefore our integral
representations (\ref{compl_pow_def}, \ref{compl_pow_def_quant}) are somewhat
complementary to the
usual ones (see, for example, the work of Varchenko and Schehtman
{}~\cite{sch_varch_1}
 for the classical case and
 of Matsuo ~\cite{mat}, Reshetikhin ~\cite{resh}
 for the quantum case ). The relation between them is unclear but, hopefully
can
be established by a special choice of a contour in (\ref{compl_pow_def},
\ref{compl_pow_def_quant})
and sending ``branching coefficients'' of $V_{i}'$s to 0;

(ii) The appearance of modules realized in multi-valued functions is forced
here by the singular
vector formula (\ref{sing_vect_form}). On the other hand, such modules have
naturally emerged in the
context of WZW model for $\widehat{\gtsl_{2}}$. They were first put forward by
 P.Furlan, A.Ch.Ganchev, R.Paunov and V.B.Petkova in \cite {f_g_p_p} in order
to get the
Virasoro minimal models correlators from the solutions for KZ equations via a
sort of a quantum
Drinfeld-Sokolov reduction. The paper  \cite {f_g_p_p} also contains some new
integral
representations for correlation functions. The relation between these integral
formulas and ours
is not clear but as P.Furlan, A.Ch.Ganchev, R.Paunov and V.B.Petkova kindly
informed us
at least some of their formulas can be recovered using ours.

 As is shown in \cite{awata},  ``the singular
vector decoupling condition'' implies that in the fractional level case a
non-zero vertex operator
takes values in a module $V(\mu,\lambda)$ for some non-integral $\mu$;

(iii) In the main body of the paper we only discuss a $q-$deformation of our
results on solutions to
the classical KZ equations and do not mention a possibility of doing the same
with regards to
intertwining operators. However this possibility does exist at least in the
case of
$U_{q}(\gtsl_{2})$.  The above mentioned modules $V(\mu,\lambda)$ admit a
deformation
$V_{q}(\mu,\lambda)$ (see sect.\ref{ExplicitFormulasorU_q(widehatgtsl_2} ). As
a linear space
$V_{q}(\mu,\lambda)$ is independent of $\lambda$ and, therefore, the operator
$E^{\beta},\;\beta\in\nc$, defined by (\ref{compl_pow_def_quant}) can be
regarded as an operator
\[E^{\beta}:V_{q}(\mu,\lambda)\rightarrow V_{q}(\mu+\beta,\lambda-2\beta).\]
Direct calculation shows that
\[E^{\lambda + 1}:V_{q}(\mu,\lambda)\rightarrow V_{q}(\mu+\lambda +
1,-\lambda-2)\]
is $U_{q}(\gtsl_{2})-$linear, which is, of course absolutely analogous to Lemma
\ref{int_sl2}. We
believe that this can also be done in a more general context and hope to
discuss it in the next paper.

{\bf Acknowlegments.} Our thanks are due to A.Kirillov, M.Jimbo, T.Miwa,
M.Kashiwara, F.Smirnov,
A.Tsuchiya, Y.Yamada
 for valuable discussions, and to K.Mimachi, who gave one of us first
lessons in $q-$analysis. Parts of the work were reported at RIMS (Kyoto
University), Nagoya and Tokyo
University and National Laboratory for
 High Energy Physics  ( KEK ) at Tsukuba. This work was done when we were
visiting RIMS and Mathematics Department of Kyoto University .

\section {{\bf Construction of Intertwining Operators.}}
\label{Construction_of_Intertwining_Operators}

\subsection {{\bf Flag Manifolds and Line Bundles }}
\label{FlagManifoldsLineBundles}

In this section we list main geometric notions related to flag manifolds and
line bundles. All the
material is pretty standard, except the notion of ``a linear bundle related to
a complex divisor'',
which might be new ( and interesting in its own right ). In order not to get
bogged down in
technicalities we choose the following strategy: firstly, we discuss everything
in a
finite-dimensional setting and then describe the necessary modifications in the
affine case.

1. Let $G$ be a complex simple finite-dimensional Lie group related to
a complex simple finite-dimensional Lie algebra $\gtg$. ( The class of such
algebras is distinguished
among all Kac-Moody algebras by the condition that the Cartan matrix $A$ is
positive definite, see
Introduction.) Denote by $N_{-},\;N_{+},\;T$ the subgroups of $G$ related to
the following
subalgebras of $\gtg$: $\gtnm,\;\gtnp,\;\gth$ (resp.).

The flag manifold $F$ is said to be $F=G/B$. It is known that $F$ is a smooth
projective variety.
Using this we accept the following terminology: by regular  function  over a
given open set we mean a
meromorphic function with singularities lying outside this set; by analytic
function we mean a
(probably multi-valued) function splitting into a power series in a
neighborhood of any non-singular
point.

$F$ possesses the standard cellular decomposition into Schubert cells
$\dot{X}_{w}$ numerated by
elements $w$ of the Weyl group $W$. To describe it recall that an alternative
way
(to the one chosen in Introduction) to define $W$ is as follows:
\[W=\, Norm(T)/T,\]
where $Norm(T)$ stands for the normalizer of $T$. Therefore $W$ is naturally
embedded in $F$ and one
sets $\dot{X}_{w}=N_{+}w$.

It is known that $dim \dot{X}_{w}=l(w)$, where $l(.)$ is the length function on
$W$.
The Schubert variety $X_{w}$ is said to be a completion  of $\dot{X}_{w}$ in
metric topology. One
shows, completing the construction of the cellular decomposition, that
\begin{equation}
\label{celldecompintw}
F=\bigcup_{w\in W}\dot{X}_{w},\;X_{w}=\bigcup_{v\leq w}\dot{X}_{v}
,
\end{equation}
where $\leq$ stands for the Bruhat ordering on $W$.

The decomposition (\ref{celldecompintw}) is in a sense attached to the point
$B\in F$. For an
arbitrary $x\in F$ set
\[X_{w}(x)=g(x)X_{w},\]
where $g(x)\in G$ satisfies $g(x)B=x\in F$. Though the last property does not
determine $g(x)$
uniquely, one verifies that $X_{w}(x)$ is independent of the choice of $g(x)$.
It follows that the
 map $x\rightarrow X_{w}(x)$ is equivariant: $X_{w}(gx)=gX_{w}(x)$.

 2. Consider the cartesian product $F\times F$. $G$ naturally acts on $F\times
F$, $G$-orbits
$\dot{Y}_{w}$ being
numerated by elements of $W$, where
  $\dot{Y}_{w}$ is said to be all  $(x,y)\in F\times F$ satisfying the
following condition: there
exists $g\in G$ ( depending on $(x,y)$)
such that $(gx,gy)\in(B,
\dot{X}_{w})$. We denote by $Y_{w}$ a completion of $\dot{Y}_{w}$ in metric
topology.

There are equivariant projections
\[pr_{i}:\;F\times F\rightarrow F,\;i=1,2\]
(on each factor) it follows from the definitions that its restriction to
$Y_{w}\subset F\times F$
is a fibration
\[pr_{i}\mid_{Y_{w}}:\,Y_{w}\rightarrow F\]
with the fiber $X_{w}(x)$ over $x\in F$. The last assertion could have been
chosen as a definition
of $X_{w}(x)$.

3. {\em The Picard group and line bundles}. A divisor on a complex variety $X$
is said to be a complex
subvariety of complex codimension 1. One may consider a free abelian group
generated by all divisors
on $X$. For example, a meromorphic function $f$
or, more generally, a meromorphic section of a line bundle determines
an element of this group, denoted by $(f)$ and called divisor of $f$, where
zeros of $f$ enter $(f)$
with $+$ and poles - with $-$.

The Picard group $Pic(X)$ is said to be the above free abelian group generated
by all divisors
on $X$ modulo subgroup of divisors of meromorphic functions. It is evident that
the map assigning to
a linear bundle $\cl$ over $X$ a divisor of  its (arbitrary) meromorphic
section, regarded as an
element of $Pic(X)$, is a 1-1 correspondence.

In the case $X=F$ we already have a collection of $n+1$ (rank of $\gtg$)
divisors: they are given
by the codimension 1 Schubert varieties $X_{i}=X_{w_{0}r_{i}},\;0\leq i\leq n$,
where $w_{0}\in W$
is an element of the maximal length and $r_{i}$ is a reflection at the $i$-th
simple root. It is
well-known that
\begin{equation}
Pic(F)=\oplus_{i=0}^{n}\nz X_{i}.
\label{isom_pic_sch}
\end{equation}

4.{\em Induced line bundles.} $F$ being a homogeneous space, affords another
construction of a line
bundle. We will be saying that a functional (weight) $\lambda\in\gth^{\ast}$ is
integral if
$\lambda(H_{i})\in\nz,\;0\leq i\leq n$. Each weight $\lambda$ determines a
character of $\gth$, for
the last algebra is commutative and each integral weight can be lifted to a
character of the group
$T$, which we also denote by $\lambda$.

Consider a projection

\[\pi:\;G/N_{+}\rightarrow F=G/B.\]

$\pi$ is a principal $T-$fibration: there is a natural right fiber-wise action
of $T$ on $G/N_{+}$,
since $T$ normalizes $N_{+}$. Denote by $\Omega^{0}_{\lambda}(G/N_{+})$ the
sheaf of regular
$\lambda-$homogeneous functions on $G/N_{+}$, i.e. functions satisfying
$f(xt)=\lambda(t)f(x),\;t\in
T$.

\begin{lemma} For an integral $\lambda\in\gth^{\ast}$ there exists a unique
line bundle $\co(\lambda)$
satisfying:

(i) $\pi^{\ast}\co(\lambda)$ is trivial;

(ii) $\pi^{\ast}$ establishes an isomorphism between the sheaf of regular
sections of $\co(\lambda)$
and $\Omega^{0}_{\lambda}(G/N_{+})$. In particular, the spaces of meromorphic
sections of
$\co(\lambda)$ and meromorphic $\lambda-$homogeneous functions on $G/N_{+}$ are
isomorphic;

(iii)the divisor of  $\co(\lambda)$ is equal to $\sum_{i}\lambda(H_{i})X_{i}\in
Pic(F)$.
\label{indlinbundl}
\end{lemma}

5. {\em The Borel-Weil theory.} The group $G$ (and, therefore, the algebra
$\gtg$) acts
by left translations on $G/N_{+}$. This action preserves $\lambda-$homogeneous
functions and
therefore, determines a structure of $\gtg-$module on meromorphic sections of
$\co(\lambda)$.

An integral weight $\lambda$ is called dominant if $\lambda(H_{i})\geq
0,\;0\leq i\leq n$.

\begin{theorem}[Borel - Weil]. Let $\lambda$ be an integral weight. Then

(i) $\Gamma (\dot{X_{w_{0}}},\, \co(\lambda))\approx M^{c}(\lambda)$, where
$M^{c}(\lambda)$ stands for a contragredient Verma module (see Introduction);

(ii) $\Gamma (F,\, \co(\lambda))=0$ unless $\lambda$ is dominant. If the latter
condition is
satisfied then $\Gamma (F,\, \co(\lambda))$ is isomorphic to the irreducible
$\gtg-$module with
highest weight $\lambda$.
\label{borelweiltheor}
\end{theorem}

6. {\em Non-integral case.} The item (i) of the Borel-Weil theorem can be
understood as a
realization of a contragredient Verma module in functions on the ``big''
Schubert cell. Really,
having a meromorphic section of $\co(\lambda)$ fixed and identified with a
constant function, one
identifies all meromorphic sections of $\co(\lambda)$ with functions on the big
cell, elements of
$\gtg$ being realized by certain order 1 differential operators. ( See, for
example, explicit
formulas for the $\gtsl_{n}-$ case in ~\cite{feig_fren_0}). Then one observes
 that the formulas depend on
$\lambda$ in a polynomial way and, therefore, make sense for an arbitrary
non-integral $\lambda$.
This completes the realization of $M^{c}(\lambda)$ in functions on the big
cell. However
in
order to construct  integral operators we would
like to keep track of  geometric nature of sections $\co(\lambda)$ for an
arbitrary $\lambda$
 (see the $\gtsl_{2}-$case in Introduction).

Observe that the  fibration $\pi:\;G/N_{+}\rightarrow F=G/B$ is trivial over
the big cell
$\dot{X}_{w_{0}}$ and we fix a trivialization $\phi$. Though $\phi$ is not
determined uniquely,
one may say that it is determined ``almost uniquely'': $\phi$ produces an
isomorphism
$\phi(\dot{X}_{w_{0}})\approx \dot{X}_{w_{0}}\times T$ and, therefore, any
other trivialization
is different from $\phi$ by an element of $T$.

{\em Definition.}
In view of Lemma \ref{indlinbundl} it is natural to say that for an arbitrary
weight $\lambda$
$\tilde{\co}(\lambda,\phi)$
is a sheaf over $\dot{X}_{w_{0}}$ defined as follows: the space of its sections
$\Gamma (U, \tilde{\co}(\lambda,\phi))$ over an open set $U\subset
\dot{X}_{w_{0}}$ is said to be the
space of functions on an open neighborhood of $\phi(U)\subset G/N_{+}$,
$\lambda-$ homogeneous and
regular on $\phi(U)$.

The above discussion  implies that for a pair of trivializations $\phi,\psi$
the corresponding sheaves are canonically isomorphic:
$\tilde{\co}(\lambda,\phi)\approx\tilde{\co}(\lambda,\psi)$.

Observe also that if $\lambda$ is integral then each $f\in \Gamma (U,
\co(\lambda))$ uniquely
determines a meromorphic section of $\co(\lambda)$ understood
 as a line bundle over $F$ with the divisor $\sum_{i}\lambda
(H_{i})$. In this sense the new definition contains the old one as a particular
case.

{\bf Convention.} Using these 2 remarks we will run the risk of reducing the
notation
$\tilde{\co}(\lambda,\phi)$ to $\co(\lambda)$.

For $f\in \Gamma (U, \co(\lambda)),\;g\in \Gamma (U, \co(\mu)),\;t\in\nc$
multiplication and
exponentiation gives rise to the following operations:
\begin{eqnarray}
f,g\;&\mapsto& fg \in \Gamma (U, \co(\lambda+\mu)),\label{multsectint}\\
f\;&\mapsto& f^{t}\in\Gamma (U, \cl\otimes\co(t\lambda))
\mbox{ if $U$ is simply connected }\label{expsectint}.
\end{eqnarray}
where $\cl$ is a sheaf of continuous branches of ( usually multi-valued )
function $f^{t}$.

Another example of dealing with $\co(\lambda)$ for an arbitrary $\lambda$ is
obtained by considering
the $SL_{2}-$case. First of all, for $SL_{2}$ one
  has: $N^{+}=\nc^{\ast}$, $SL_{2}/N^{+}\approx \nc^{2}-\{(0,0)\}$ ( see
{}~\cite{b_g_g}),
 $F= \cp^{1}$, the big cell is $\nc$ and $\pi$ over $\nc$ is simply the
projection $\nc\times
\nc^{\ast}\,\rightarrow \nc$. A weight is a complex number:
 $\lambda\in\nc$ and, therefore sections of $\co(\lambda)$ are naturally
determined by
expressions $f(x)\,dx^{-\lambda/2}$, where $f(x)$ is a meromorphic function.
Looking at Introduction,
formula (\ref{fam_sl2_mod}) one realizes that the integral
operator for $\gtsl_{2}$ (see (\ref{int_sl2_form})) was obtained by making use
of the operations
(\ref{multsectint},\ref{expsectint}).

Further
for an arbitrary $G$ and a fixed simple root $\alpha_{i}$ of $G$ one may
consider an inclusion
$SL_{2}\subset G$, on the level of Lie algebras  determined by the
$sl_{2}-$triple:
$\langle E_{i},H_{i},F_{i}\rangle$. This gives rise to the inclusion of a flag
manifold $F_{i}\subset
F$. One realizes that the latter identifies $F_{i}$ with the Schubert variety
$X_{r_{i}}$. The
following assertion ( of course well-known for an integral
$\lambda\in\gth^{\ast}$ )
 now becomes a tautology:
\begin{equation}
\label{restrlinbundlintw}
\co(\lambda)\mid_{X_{r_{i}}}=\co(\lambda(H_{i})).
\end{equation}

{\bf Remark.} The necessity to understand geometric meaning of a ``complex
power of a line bundle''
appeared several years ago in connection with the developments in the string
theory. The matter was
treated in \cite{beil_sch},\cite{voron} in the framework of complex curves
rather than homogeneous
spaces as it is here.

7. {\em The affine Lie group case.} Everything we have encountered with so far
directly generalizes to the case of an affine Lie group. Making a reference to
the book
{}~\cite{pr_seg} (  an exposition of this theory in even more general setting
can be found in
{}~\cite{kum} ),
 we here
restrict ourselves to a few comments.

An affine Lie group
 modulo center is just a loop group $L(G)$, where an element of the latter is
understood
as a smooth map $\gamma:\,S^{1}\rightarrow G$. The corresponding Lie algebra -
an affine Lie algebra
- is included into the class of Kac-Moody algebras (see  sect.
\ref{tr_f_k_z_eq}).
 A Borel subgroup $\tilde{B}$
and a ``maximal nilpotent subgroup'' $\tilde{N}^{\pm}$
are said to consist of
boundary values of analytic inside the unit disc maps $\gamma:\,\{z:\,|z|\leq
1\}\rightarrow G$,
 satisfying $\gamma (0)\in B$
($\gamma (0)\in N^{\pm}$ resp.). The
flag manifold is said to be the quotient $L(G)/\tilde{B}$. It also possesses a
cellular decomposition
but now one has to distinguish between cells of finite dimension and finite
codimension. The former
are orbits of elements of an affine Weyl group under $\tilde{N}^{+}$, while the
latter are also with
respect to $\tilde{N}^{-}$. (This could have been also said with regards to the
finite-dimensional
case, producing equivalent description of the cells.)

It is crucial that Borel-Weil theory (Theorem \ref{borelweiltheor} ) works in
the affine case as well
(see the above-mentioned book ~\cite{pr_seg} and ~\cite{kum,feig_fren}).
 One also realizes that the correspondence among
codimension 1 Schubert varieties, line bundles and induced line bundles,
 our discussion of the fibration
$\pi:\,G/N^{+}\,\rightarrow G/B$ together with the subsequent definition of the
sheaves
$\co(\lambda)$ is  independent of the passage to the affine case.

\subsection { Construction of an Intertwining Operator related to a simple
root.}
\label {Construction_of_an_Intertwining_Operator_related_to_a_simple root}
In this section we use the following unified notations:
$\gtg,\gtb,...,G,B,F...$ stand for either
simple finite-dimensional or affine Lie algebra, its Borel subalgebra,..., the
corresponding Lie
group, its Borel subgroup, the corresponding flag manifold...(resp.),
$\dot{X}_{i}=N^{-}r_{i}\,0\leq i\leq n$ stands for the Schubert cell of
codimension 1,
 $\dot{X}_{w}, X_{w},
\dot{Y}_{w}, Y_{w},\, w\in W$ usually
 stand for a Schubert cell (variety, $G$-orbit in $F\times F$...resp.)
of a finite dimension $l(w)$, except $\dot{X}_{w_{0}}$ which stands for the big
Schubert cell. The
last convention is consistent with the previous ones in the finite-dimensional
case, while it is not
quite so in the affine case, for in the latter case there is no element of the
maximal length in the
Weyl group.

As is known, for example from the representation theory of complex simple Lie
groups,
the  scheme of constructing intertwiners as integral operators could be as
follows. Consider
the direct product $F\times F$ and 2 projections $pr_{1},\;pr_{2}$ on the 1st
and the 2nd factors
respectively. Take a section $f\in\co (\lambda)$ pull it back on $F\times F$ to
get
$pr_{1}^{\ast}f\in pr_{1}^{\ast}(\co (\lambda))$, ``pair'' it with
``something'' defined on $F\times
F$ (or on a $G$-invariant subset of $F\times F$)
, ``integrate the result over the fibres of $pr_{2}$'' and so get a section of
a bundle defined over
$F$. We realize this scheme as follows.

{\bf A. Integral case.} Let $\lambda\in \gth^{\ast}$ be integral.
Let $D\subset F\times F$ be the diagonal and for a pair of line bundles
$\ch_{1},\ch_{2}$ over $F$
denote by $\ch_{1}\odot\ch_{2}$ their exterior product as a line bundle over
$F\times F$.
One observes that the line bundle $(\co(\alpha_{i})\odot
\co(\alpha_{i}))\mid_{Y_{r_{i}}}$
is related to the divisor $2D$ ( $D$ is understood here as a divisor inside
$Y_{r_{i}}$ )
and we pick up a  section
$s_{i}$ of $(\co(\alpha_{i})\odot
\co(\alpha_{i}))\mid_{Y_{r_{i}}}$ so that $(s_{i})=2D$. It is important that
$s_{i}$ is $G-$invariant
(since $D$ is also).

Another observation is that if one sets
\[\ca_{i}(\lambda)
=((\co(\lambda)\odot\nc)\otimes (\co(\alpha_{i})^{-\frac{1}{2}(\lambda
+2\rho)(H_{i})}\odot
(\co(\alpha_{i})^{-\frac{1}{2}(\lambda +2\rho)(H_{i})}))\mid_{Y_{r_{i}}},\]
then for any $f\in\Gamma (X^{0}_{w_{0}},\co (\lambda))$,
 $pr_{1}^{\ast}f\otimes s_{i}^{-\frac{1}{2}(\lambda +2\rho)(h_{i})}$
 is a (meromorphic) section of $\ca_{i}(\lambda)$.

\begin{lemma}
Restriction of $\ca_{i}$ to $pr_{2}^{-1}(x)\cap Y_{r_{i}}$ ($\approx \cp ^{1}$)
is a sheaf of
differential forms.
\label {restr_lemma_int}
\end{lemma}
{\bf Proof.} The following property of a sheaf $\co(\mu)$ on a flag manifold (
see
(\ref{restrlinbundlintw}) )
 \[deg (\co(\mu)\mid_{X_{r_{i}}(x)})=\mu(H_{i})\]
implies that for any (meromorphic) section $f$ of $\co(\mu)$ and the
above-defined $s_{i}$
and for an arbitrary coordinate $t$ on
$pr_{2}^{-1}(x)\cap Y_{r_{i}}$ ($\approx X_{r_{i}}(x)$) one has
\[pr_{1}^{\ast}f\mid_{X_{r_{i}}(x)}=p(t)dt^{-\lambda(H_{i})/2},\]
\[s_{i}\mid_{X_{r_{i}}(x)}=C(t-x)^{2}dt^{-1},\]
where $p(t)$ is a rational
function of $t$ and $C$ is independent of $t$. It follows that the restriction
of
$pr_{1}^{\ast}f\otimes
 s_{i}^{-\frac{1}{2}(\lambda+2\rho)(H_{i})}$ to $X_{r_{i}}(x)$ is a
differential
form. $\Box$

Making use of this lemma
 we denote by

\begin{equation}
\label{def_of_pair}
\int_{pr^{-1}_{2}(y)}pr_{1}^{\ast}f\otimes s_{i}^{-\frac{1}{2}(\lambda
+2\rho)(h_{i})}
\end{equation}

the residue of $pr_{1}^{\ast}f\otimes s_{i}^{-\frac{1}{2}(\lambda
+2\rho)(H_{i})}\mid_{pr^{-1}_{2}(y)}$ at the point $(y,y)$
for
any $f\in\Gamma (\dot{X}_{w_{0}},\co (\lambda))$ and $y\in F$.

It follows from the definition that
$\int_{pr^{-1}_{2}(y)}pr_{1}^{\ast}f\otimes s_{i}^{-\frac{1}{2}(\lambda
+2\rho)(h_{i})}$
 is a section of
the bundle $\ca_{i}\otimes (\co(\alpha_{i})\odot\nc)\mid_{Y_{r_{i}}}$. The last
bundle can be
restricted to the diagonal and then pushed forward on $F$ via
the isomorphism $pr_{2}\mid_{D}:\;D\approx
F$. The result is the bundle $\co(\lambda)\otimes\co(\alpha_{i})^{-(\lambda
+\rho)(H_{i})}=\co(r_{i}\cdot\lambda)$.
  Denote by
$T_{i}(f)$ the image of
$\int_{pr^{-1}_{2}(y)}pr_{1}^{\ast}f\otimes s_{i}^{-\frac{1}{2}(\lambda
+2\rho)(h_{i})}$ under
this composition map. Thus we have constructed a $\nc-$linear map
\[T_{i}:\;\Gamma (\dot{X}_{w_{0}},\co(\lambda))\rightarrow\Gamma
(\dot{X}_{w_{0}},\co(r_{i}\cdot\lambda)).\]
\begin {theorem} $T_{i}$ is a morphism of $\gtg-$modules.
\label{int_case}
\end{theorem}
{\bf Proof} immediately follows from the construction. $T_{i}$ was defined to
be a composition of
the following operations on sections of $\gtg-$bundles:

(i) pulling back from $F$ to $F\times F$ via $pr_{1}$,

(ii) tensoring with a $\gtg-$invariant section $s_{i}^{-\frac{1}{2}(\lambda
+2\rho)(h_{i})}$
of the line
bundle $(\co(\alpha_{i})^{-\frac{1}{2}(\lambda +2\rho)(H_{i})}\odot
(\co(\alpha_{i})^{-\frac{1}{2}(\lambda +2\rho)(H_{i})})\mid_{Y_{r_{i}}}$,

(iii) pushing down on $F$ via $pr_{2}$.

Each of these operations is obviously $\gtg -$ linear.$\Box$

{\bf Remarks.} 1. One easily realizes that the constructed morphism $T_{i}$ of
contragredient Verma
modules is dual to the embedding of Verma modules determined by the singular
vector $F_{i}^{\lambda
(H_{i})+1}v_{\lambda}$ (see (\ref{simplestsingvectorintr}) ). On the other
hand, the last morphism may
not exist, though our $T_{i}$ is still defined. In this case $\lambda
(H_{i})+1< 0$, the integrand in
the definition of $T_{i}$ is regular (see also (\ref{def_of_pair}) ) and,
therefore,
$T_{i}=0$ as one should have expected.

2. One may want to define an intertwiner
\[T_{w}:\Gamma (\dot{X}_{w_{0}},\co(\lambda))\rightarrow\Gamma
(\dot{X}_{w_{0}},\co(w\cdot\lambda))\]
related to  an arbitrary element of the Weyl group
$w=\cdots r_{i_{2}}r_{i_{1}}$ as a composition of the constructed $T_{i}$'s:
$T_{w}=\cdots T_{i_{2}}T_{i_{1}}$. However this does not work since by the
above remark  one
of the factors is usually 0. To obtain the desired formula one has to ``step
aside'' and
 define $T_{i}$'s
for a suitable module related to a generic weight $\lambda$.

{\bf B. Nonintegral case.} In
order to construct an analogue of $T_{i}$ for an arbitrary ``highest weight''
we first define a
suitable

{\em  1. Family of $\gtg-$modules. } Fix some group element $g\in G$ and
denote by $X^{opp}_{0},\ldots ,X^{opp}_{n}$ the ``opposite''
Schubert cells defined by: $X^{opp}_{i}=gX_{i}$. If $dim\gtg<\infty$, then the
natural choice is
$g=w_{0}$, where $w_{0}\in W$ is the element of the maximal length, which
explains the terminology.

 The space $F^{0}
=F-\bigcup _{0\leq j\leq n}X_{j}- \bigcup _{0\leq j\leq n}X^{opp}_{j}$
 will serve as an analogue of $\nc^{\ast}$ in the
$\gtsl_{2}-$case.

 For any local
system
$\cl$ over $F^{0}$ and any $\lambda\in\gth^{\ast}$ we set $\co (\cl ,\lambda)
=\cl\otimes(\co(\lambda)\mid_{F^{0}})$. The algebra $\gtg$ naturally acts on
sections  of
$\co (\cl ,\lambda)$ since it acts on sections of both $\cl$ and
$\co(\lambda)$. To each local
system $\cl$ we associate a vector function (collection of exponents )
\[\mu=(\mu_{0},\ldots,\mu_{n}):\;F^{0}\rightarrow \nc^{n+1}\]
as follows. One proves that the intersection number of $X_{r_{i}}(x)$ with
$X_{j}$ as well as
with $X_{j}^{opp}$ is equal to 1 if $i=j$ and - to 0 otherwise
for  all $x\in F^{0}$ ( this can be derived from the Borel-Weil theorem
\ref{borelweiltheor} and
(\ref{restrlinbundlintw}) ).
 To each $x\in F^{0}$ and a number $i$ we  associate a projective line
$X_{r_{i}}(x)$ with 2
marked points: $t_{0}(x)=X_{r_{i}}(x)\cap
X_{i},\;t_{\infty}(x)=X_{r_{i}}(x)\cap X_{i}^{opp}$ . The
projective line punctured at 2 points is homotopically equivalent to a circle .
 Therefore, the restriction of $\cl$ to $X_{r_{i}}(x)$  determines (and is
determined by )  a complex
number $\mu_{i}(x)$ - its monodromy coefficient.

\begin{example} It follows from the definition that the divisors
$(X_{j}),(X_{j}^{opp})
,\;0\leq j\leq n$ are
equivalent (they determine the same line bundle, namely  $\co(\lambda^{(j}))$
where
$\lambda^{(j)}$ is the
$j-$th fundamental weight). We fix functions $\phi_{0},\ldots ,\phi_{n}$ on $F$
so that
$(\phi_{j})=(X_{j})-(X_{j}^{opp})$. For a complex vector $
\mu=(\mu_{0},\ldots,\mu_{n})$ define a multivalued function $\phi=\prod_{1\leq
j\leq
n}\phi_{j}^{\mu_{j}}$. Singlevalued branches of this function determine a local
system $\cl_{\mu}$ on
$F^{0}$, the vector function $
\mu=(\mu_{0},\ldots,\mu_{n})$ being constant in this case. The bundle
$\co(\cl_{\mu},\lambda)$
 is the simplest example of bundles we will be dealing with.
\label{ex_of_loc_syst}
\end{example}

 We will  construct  an intertwining operator $T_{i}$ taking a section of
$\co(\cl ,\lambda)$ to
a section of $\co(\cm ,r_{i}\cdot\lambda)$ for arbitrary $\cl$ and $\lambda$,
$\cm$ being uniquely
determined by $\cl$ and $\lambda$. Moreover it will be given by essentially the
same formula as in the
integral case.Set
\[\ca_{i}=((\co(
\cl ,\lambda)\odot\nc)\otimes (\co(\alpha_{i})^{-\frac{1}{2}(\lambda
+2\rho)(H_{i})}\odot
(\co(\alpha_{i})^{-\frac{1}{2}(\lambda +2\rho)(H_{i})}))\mid_{Y_{r_{i}}}.\]
(See the previous section, especially
(\ref{multsectint},\ref{expsectint}) ,
 for the discussion of complex powers of line bundles on flag manifolds.)

As before, the choice of the exponent $-\frac{1}{2}(\lambda +2\rho)(H_{i})$
implies that
the restriction of $\ca_{i}$ to any fiber of $pr_{2}$ $\approx \cp^{1}$ is
``tensorially'' a sheaf
differential forms, more precisely this is a sheaf of differential forms on
$\cp^{1}$ with
coefficients in a certain local system. We now construct this local system.

2.
 Each projective line $pr^{-1}_{2}(x)\approx X_{r_{i}}(x)$ is equipped with
three marked points $t_{0}(x),\; t_{\infty}(x),\; x$. Projective line punctured
at 3 points is
homotopically equivalent to the union  of 2
circles.
The 1st homotopy group  $\pi_{1}(X_{r_{i}}(x)-\{t_{0}(x),t_{\infty}(x),x\})$ is
a free group on
2 generators $\sigma_{1}(x),\sigma_{2}(x)$ which can be thought of as
a pair of circles enclosing
the points $t_{0}(x)$ and $x$ respectively. Denote by $\cl_{i}(\cl,\lambda;x )$
the local system on
$X_{r_{i}}(x)-\{t_{0}(x),t_{\infty}(x),x\}$
determined by the following 1-dimensional representation of
$\pi_{1}(X_{r_{i}}(x)-\{t_{0}(x),t_{\infty}(x),x\})$:
\[\sigma_{1}(x)\mapsto exp(-2\pi\sqrt{-1}\mu_{i}),\;
\sigma_{2}(x)\mapsto \exp(2\pi\sqrt{-1}\lambda_{i}(H_{i})).\]
(Recall that $\{\mu_{i}\}$ is the set of exponents.)

For generic $\mu,\lambda$ the 1st homology group
$H_{1}(X_{r_{i}}(x)-\{t_{0}(x),t_{\infty}(x),x\},\;\cl_{i}(\cl,\lambda;x )$
is 1-dimensional and is
generated by the cycle
$\sigma_{1}(x)\sigma_{2}(x)\sigma_{1}^{-1}(x)\sigma_{2}^{-1}(x)$. Now the
relation of $\ca_{i}$ to
the constructed local system is as follows.
We again pick out a $G-$invariant section $s_{i}$ of the bundle
$(\co(\alpha_{i})\odot
(\co(\alpha_{i})))\mid_{Y_{r_{i}}}$ such that $(s)=2D$.
\begin{lemma}
\label{restr_lemma_nonint}
For any $f\in\Gamma (\dot{X}_{w_{0}},\co (\cl,\;\lambda))$ the restriction of
$pr_{1}^{\ast}f\otimes s_{i}^{-\frac{1}{2}(\lambda +2\rho)(H_{i})}$
to a fiber $pr_{2}^{-1}(x)$ represents a cohomology class of
$X_{r_{i}}(x)-\{t_{0}(x),t_{\infty}(x),x\}$ with coefficients in
$\cl_{i}(\cl,\lambda;x )$.
\end{lemma}
{\bf Proof.} As in Lemma \ref{restr_lemma_int}  the restriction of
$pr_{1}^{\ast}f\otimes s_{i}^{-\frac{1}{2}(\lambda +2\rho)(H_{i})}
$ to $X_{r_{i}}(x)$ for a fixed $x$ is given by the formula
\[\frac{f(t)}{(t-x)^{\lambda (H_{i})+2}}dt,\;f(t)\in t^{\mu_{i}}\nc
[t,t^{-}],\]
where with a certain abuse of notation $x$ stands for the value of $t$ at the
point $x$ (c.f.
formula (\ref{compl_der})). $\Box$

We are almost ready to define an intertwining operator by the formula literally
coinciding with that
of Theorem\ref{int_case}
 but in order to understand what its image is we have to introduce one more
local  system on $F^{0}$.

 3. The assignment $F^{0}\ni x\mapsto
X_{r_{i}}(x)-\{t_{0}(x),t_{\infty}(x),x\}$
naturally determines a
fibration over $F^{0}$ with the fiber: projective line punctured at 3 points.
This fibration gives
rise to a line bundle with a fiber over $x\in F^{0}$ equal to
 $H_{1}(X_{r_{i}}(x)-\{t_{0}(x),t_{\infty}(x),x\},\;\cl_{i}(\cl,\lambda;x ))$.
Denote this line bundle by
$\cl_{i}(\cl,\lambda)$. A trivialization of the fibration with the fiber
projective line punctured
at 3 points over a disk induces a trivialization of  $\cl_{i}(\cl,\lambda)$
over the same disk and,
therefore an identification of
$H_{1}(X_{r_{i}}(x)-\{t_{0}(x),t_{\infty}(x),x\},\;\cl_{i}(\cl,\lambda;x )$
for all $x$ from this disk. Obviously the last identification is independent of
a trivialization and
gives rise to a flat connection called the Gauss-Manin connection. There arises
a local system of horizontal
sections of $\cl_{i}(\cl,\lambda)$ with respect to this connection.

4.{\em Formula for an intertwiner.} We, firstly, have to fix an arbitrary
simply connected open
subset $\tilde{F}^{0}\subset F^{0}$, for non-trivial sections of our sheaves
may not exist over
non-simply connected open sets.
Now define a linear map

\[\tilde{T}_{i}:\;\Gamma (\tilde{F}^{0},\cl_{i}(\cl ,\lambda)\otimes \co(\cl
,\lambda))\rightarrow
\Gamma(\tilde{F}^{0},\co(\cl, r_{i}\cdot\lambda))\]
 to be a composition of the following operations:

(i) $pr_{1}^{\ast}\otimes pr_{1}^{\ast}$: it takes a  section
 $\sigma\otimes f$ of $\cl_{i}(\mu,\lambda)\otimes\co(\mu,\lambda)$
  into a section $pr_{1}^{\ast}\sigma\otimes pr_{1}^{\ast}f$
and  of the corresponding sheave over $Y_{r_{i}}$;

(ii)  $\otimes s_{i}^{-\frac{1}{2}(\lambda + 2\rho)(H_{i})}$: it tensors the
result of the previous
operation by $s_{i}^{-\frac{1}{2}(\lambda + 2\rho)(H_{i})}$; the result of the
composition map
$(\otimes s_{i}^{-\frac{1}{2}(\lambda + 2\rho)(H_{i})})\circ
(pr_{1}^{\ast}\otimes pr_{1}^{\ast})$
is a section of the bundle $pr_{1}^{\ast}(\cl_{i}(\cl,\lambda))\otimes\ca_{i}$;

(iii) $\int_{pr_{2}^{-1}}$: by Lemma\ref{restr_lemma_nonint} there is a natural
pairing of
sections of $pr_{1}^{\ast}(\cl_{i}(\cl,\lambda))$ with those of $\ca_{i}$ which
for a fixed point
$(x,y)\in F^{0}\times F^{0}$ is nothing but the value of the cohomology class
determined by the
restriction of the section of $\ca_{i}$ to the fiber $pr_{2}^{-1}(y)\cap
Y_{r_{i}}$ on the homology
class determined by the value of the
 section of $pr_{1}^{\ast}(\cl_{i}(\cl,\lambda))$ at the point $(x,y)$; the
result is a section of a
certain bundle which can be easily calculated (see below);

(iv) $pr_{2}^{!}$: this is a composition of the restricting to the diagonal
$D\in F\times F$ and
the consecutive pushing forward on $\tilde{F}^{0}$ by means of the isomorphism
$pr_{2}:D\rightarrow
F$.

We set
\[\tilde{T}_{i}=pr_{2}^{!} \circ ( \int_{pr_{2}^{-1}}) \circ
( \otimes s_{i}^{-\frac{1}{2}(\lambda + 2\rho)(H_{i})}) \circ
( pr_{1}^{\ast}\otimes pr_{1}^{\ast}).\]
It follows from the definition (and can be seen exactly as in the integral
case)
that the constructed map acts as
\[\tilde{T}_{i}:\;\Gamma(\tilde{F}^{0},\cl_{i}(\cl ,\lambda)\otimes \co(\cl
,\lambda))\rightarrow
\Gamma(\tilde{F}^{0},\co(\cl, r_{i}\cdot\lambda)).\]
Tensoring both sides of the last equality by $\cl_{i}^{\ast}(\cl ,\lambda)$ we
finally obtain the
desired map
\[T_{i}:\;\Gamma(\tilde{F}^{0}, \co(\mu ,\lambda))\rightarrow
\Gamma(\tilde{F}^{0},\cl_{i}^{\ast}(\mu ,\lambda)\otimes\co(r_{i}\cdot\lambda))
=\Gamma(\tilde{F}^{0},\co(\cl_{i}^{\ast}(\mu ,\lambda)\otimes\cl
,r_{i}\cdot\lambda)).\]

\begin {theorem} $T_{i}$ is a morphism of $\gtg-$modules.
\label{nonint_case}
\end{theorem}
{\bf Proof} is a literal repetition of that of Theorem\ref{int_case}:
the map $T_{i}$ was defined to be a composition of maps each of them being
transparently $\gtg-$invariant
$\Box$.

{\bf Remark.} It is easy ( and might be instructive ) to examine the
$\gtsl_{2}-$ case, considered in
Introduction from this general point of view. One has:

 $V(\mu ,\lambda)\approx \Gamma(\tilde{F}^{0},\,\co (\cl_{\mu}, \lambda))$ for
an arbitrary simply
connected open $\tilde{F}^{0}\subset \nc^{\ast}$;

\[s_{i}=\frac{dx\, dt}{(t-x)^{2}};\]

$T_{i}$ from Theorem \ref{nonint_case} coincides with $T$ from Lemma
\ref{int_sl2}.

\subsection{{\bf Construction of  Intertwining Operators related to an
arbitrary element of the
Weyl group.}}
1. The simplest way to construct an intertwiner $T_{w}$ related to an element
$w\in W$ is to
consider a reduced decomposition $w=r_{i_{l}}\cdots r_{i_{1}}$ and then set
\begin{equation}
\label {simple_case_intertw}
T_{w}=T_{i_{1}}\circ T_{i_{2}}\circ\cdots\circ T_{i_{l} }.
\end{equation}
Theorem\ref{nonint_case} immediately gives
\begin{lemma} Operator $T_{w}$ as defined in (\ref{simple_case_intertw}) is a
$\gtg-$linear map
\[T_{w}:\;\Gamma^{0}(\tilde{F}^{0},\co (\cl,\lambda))\rightarrow
\Gamma^{0}(\tilde{F}^{0},\co (\cm,w^{-1}\cdot\lambda)),\]
where $\cm$ is a local system defined as follows: if one sets $\cm_{0}=\cl$,
$\cm_{j}=\cl_{i_{k-j+1}}^{\ast}(\cm_{j-1},r_{i_{k-j+1}}\cdots
r_{i{k}}\cdot\lambda)\otimes\cm_{j-1}$,
then $\cm=\cm_{k}$.
\end{lemma}

The construction implies (also see below) that each $T_{i}$ basically is an
integral of a ``1-form''
over a ``1-cycle'' and, therefore, $T_{w}$ from (\ref{simple_case_intertw}) is
an integral of a
``$k-$form'' over a  ``$k-$cycle
''. We are now going to analyze this situation and produce this ``$k-$form''
and the homology group
this ``$k-$cycle'' comes from.

2. Consider a $\cp^{1}-$fibration ( a fibration with the fiber $\cp^{1}$ )
\[\pi : A_{1}\rightarrow A_{0}.\]
For any line bundle over $A_{1}$ its degree is said to be the degree of its
restriction to (any)
fiber of $\pi$. Let $\cm_{1}$ be some line bundle over $
A_{1}$ and denote by $\co(2)_{1}$ the line bundle of vector fields over $A_{1}$
tangent to fibers of
$\pi$. Clearly the degree of $\co(2)_{1}$ is equal to 2.
Denote the degree of $\cm_{1}$ by $d_{1}$. Then
the restriction of the bundle $\cm_{1}\otimes \co(2)_{1}^{-d_{1}/2}$
to the fibers of $\pi$ is trivial and as is
well-known there is a bundle, say $\cm_{0}$, such that
\[\cm_{1}\otimes \co(2)_{1}^{-d_{1}/2}\approx \pi^{\ast}\cm_{0},\]
or
\[\cm_{1}\otimes \co(2)^{-d_{1}/2-1}\approx \pi^{\ast}\cm_{0}\otimes
\co(2)_{1}^{-1}.\]

In other words, there arises a morphism of sections
\[\phi_{1}:\;\Gamma(\pi^{-1}U,\,\cm_{1}\otimes
\co(2)_{1}^{-d_{1}/2-1})\rightarrow
\Gamma(U,\, \cm_{0}\otimes \cv_{1}),\]
where $\cv_{1}$ is a sheaf over $A_{0}$ is defined to associate to an open set
$U\subset A_{0}$ all
holomorphic sections of $\co(2)_{1}^{-1}$ over $\pi^{-1}(U)$ or, equivalently,
is a sheaf of volume
forms over the fibers of the fibration $\pi$.
  In particular, for a fixed section
$s_{1}$ of $\co(2)_{1}^{-d_{1}/2-1})$ there arises a map

\begin{equation}
\label{const_elem_1}
\phi (s_{1}):\;\Gamma(\pi^{-1}U,\,\cm_{1})\rightarrow
\Gamma(U,\, \cm_{0}\otimes \cv_{1}).
\end{equation}

 Now consider a sequence of $\cp^{1}-$fibrations
\[A_{k}\stackrel{\pi_{k}}{\rightarrow} A_{k-1}
\stackrel{\pi_{k-1}}{\rightarrow}\cdots\stackrel{\pi_{2}}{\rightarrow}A_{1}
\stackrel{\pi_{1}}{\rightarrow}A_{0}\]

together with the following data:

$\co(2)_{j}$ is a line bundle of vector fields tangent to fibers of $\pi_{j}$;

$\cm_{j}$ is a line bundle over $A_{j}$ of degree
$d_{j}$ such that
\[\cm_{j}\otimes \co(2)_{j}^{-d_{j}/2}\approx \pi^{\ast}\cm_{j-1},\;1\leq j\leq
k.\]
(The entire sequence $\{\cm_{j},\;1\leq j\leq k\}$ is uniquely determined by a
choice of the bundle
$\cm_{k}$.)

Fix sections $s_{j}$ of the bundles $\co(2)_{j}^{-d_{j}/2-1}$. Then the
repeated use of the previous
construction gives the map

\begin{equation}
\label{const_elem_2}
\phi(s_{1})\circ\cdots\circ\phi(s_{k}):\;\Gamma
(\pi_{k}^{-1}\circ\cdots\circ\pi_{1}^{-1}U,\,\cm_{k})\rightarrow
 \Gamma(U,\, \cm_{0}\otimes
\cv_{k}),
\end{equation}

where $\cv_{k}$ is a sheaf over $A_{0}$ of (fiber-wise) complex-analytic volume
forms with respect
to the composition fibration
\[\pi(s_{1})\circ\cdots\circ\pi(s_{k}):\;A_{k}\rightarrow A_{0}.\]

3. The above considerations are related to the intertwiners $T_{r_{i}}$ as
follows. First of all, the
fibration
\[Y_{i}\rightarrow F\]
can be regarded as a $\cp^{1}$-fibration with the fiber over $x\in F$ equal to
$X_{i}(x)$. Then one
easily realizes that in the flag manifold case one can easily deal with complex
powers of line
bundles. In particular, if $\co(\cl,\lambda)_{1}$ is the lift of
$\co(\cl,\lambda)$ to $Y_{i}$ then
(\ref{const_elem_1}) is valid in the form

\begin{equation}
\label{const_elem_next_2}
\phi (s_{i}(\lambda)):\;\Gamma(\tilde{F}^{0}, \co(\cl,\lambda)_{1})\rightarrow
\Gamma(\tilde{F}^{0},\co(r_{i}\cdot\lambda)_{1}\otimes \cv_{1}),
\end{equation}

where  we denote by $s_{i}(\lambda)$ the section
$s_{i}^{-\frac{1}{2}(\lambda+2\rho)(H_{i})}$
appearing, for example, in Lemma \ref{restr_lemma_nonint} and

$\cv_{1}$ is understood to be a sheaf with the fiber over $x\in F^{0}$:
differential forms over
$X_{i}(x)$ with coefficients in $\cl_{i}(\cl,\lambda;x)$. Tensoring with
$\cl_{i}(\cl,\lambda)$ and
consecutive integration gives the operator $T_{i}$.

4. We are now going to apply (\ref{const_elem_2}) to construct intertwiners
related to an arbitrary
element of the Weyl group. The possibility of doing this is based on the
following
Bott-Samelson-Demazure construction (see ~\cite{demaz} and also \cite{kum} for
the Demazure theory
in  an infinite-dimensional setting ).

 Obviously $Y_{i}$ is the set of pairs $\{(x_{0},x_{1}):\;x_{0}\in F,\;x_{1}\in
X_{i}(x_{0})\}$.
For a reduced decomposition $w=r_{i_{k}}\cdots r_{i_{1}}$ set
\[\tilde{Y}_{w}=
\{(x_{0},x_{1},\ldots,x_{k}):\;x_{0}\in F,\;x_{j}\in
X_{i_{j-1}}(x_{j-1}),\;1\leq j\leq k\}.\]

Set $w_{j}=r_{i_{j}}\cdots r_{i_{1}}$. There arises the sequence of
$\cp^{1}-$fibrations
\[\sigma_{j}:\;\tilde{Y}_{w_{j}}\rightarrow \tilde{Y}_{w_{j-1}},\]
\[\sigma_{j}:\;(x_{0},\ldots x_{j-1},x_{j})\mapsto (x_{0},\ldots x_{j-1}).\]

\begin{theorem} (see ~\cite{demaz},\cite{kum}) There exist $G-$invariant
rational equivalencies
\[\phi_{j}:\;Y_{w_{j}}\rightarrow\tilde{Y}_{w_{j}}\]
commuting with inclusions $Y_{w_{j-1}}\subset Y_{w_{j}}$ and with singularities
concentrated on
$Y_{v},\;v\leq w_{j-1}$.
\label{bott_sam_dem}
\end{theorem}

 Denote by $(\co(\alpha_{i})\odot\co(\alpha_{i}))_{j}$ the lift  of
 $\co(\alpha_{i})\odot\co(\alpha_{i})$ on
$\tilde{Y}_{w_{j}}$ via the composition map
\[\tilde{Y}_{w_{j}}\stackrel{\sigma_{j}}{\rightarrow}\tilde{Y}_{w_{j-1}}
\stackrel{\sigma_{j-1}}{\rightarrow}\cdots\stackrel{\sigma_{2}}{\rightarrow}
\tilde{Y}_{w_{1}}
\stackrel{\phi_{1}^{-1}}{\rightarrow}Y_{i_{1}}\]
and by $\co(\cl,\lambda)_{j}
$ the lift of
$\co(\cl,\lambda)
$ via the composition map
\[\tilde{Y}_{w_{j}}\stackrel{\sigma_{j}}{\rightarrow}\tilde{Y}_{w_{j-1}}
\stackrel{\sigma_{j-1}}{\rightarrow}\cdots\stackrel{\sigma_{2}}{\rightarrow}
\tilde{Y}_{w_{1}}
\stackrel{\sigma_{1}^{-1}}{\rightarrow}Y_{i_{1}}\rightarrow F .\]

Let $s_{i}^{(j)}(\lambda)$ be the lift of the section $s_{i}(\lambda)$
appearing in
(\ref{const_elem_next_2}) on $\tilde{Y}_{w_{j}}$. The same reasoning as in the
``integral'' case
and the use of (\ref{const_elem_next_2})
shows that there arises a map
$\phi(s_{i}^{(j)})$ which takes a section of $\co(\cl,\lambda)_{j}$ into a
section of
$\co(\cl,\lambda)_{j-1}$ with coefficients in the sheaf with the fiber over
$(x_{0},\ldots
,x_{j-1})$: differential forms over  $\tilde{X}_{r_{j}}(x_{j-1})$ with
coefficients in
$\cl_{i_{j}}(\cl,\lambda;x_{j-1})$. This map is $\gtg-$invariant since it is
constructed via the
$\gtg-$invariant section $s_{i}^{(j)}(\lambda)$.

Now consider the composition

\[\phi(s_{i_{1}})\circ\cdots\circ\phi(s_{i_{l}}).\]

It acts as
\begin{equation}
\phi(s_{i_{1}})\circ\cdots\circ\phi(s_{i_{l}}):\,
\Gamma(\tilde{F}^{0},\,\co(\cl,\lambda))\rightarrow\Gamma(\tilde{F}^{0},\,\co
(\cl ,
r_{i_{1}}\cdots r_{i_{l}}\cdot\lambda)\otimes Vol)
\label{almost_final_2}
\end{equation}

for a
certain sheaf $Vol$. The latter sheaf can be described as follows.

  Consider the fibration
\[\sigma_{1}\circ\cdots\circ\sigma_{k}: \tilde{Y}_{w}\rightarrow F.\]
Its fiber over $x\in F$ is  $\tilde{X}_{w}(x)$.
Now recall that each successive application of $\phi(s_{i_{j}})^{(j)}$ results
in a section of
$\co(r_{i_{k-j+1}}\cdots r_{i_{k}}\cdot\lambda)$ with coefficients in a sheaf
of differential forms
over fibers with coefficients in the sheaf obtained at the previous step. This
means that
$Vol$ is a sheaf over $F$
with the fiber
over a point $x\in F^{0}$: volume forms over $\tilde{X}_{w}(x)$ with
coefficients in a certain sheaf
over $\tilde{X}_{w}(x)$. Denote the latter sheaf by $\cv_{x}$. Unfortunately we
do not
possess  its  direct description. Nevertheless it follows from the definition
of
$\phi(s_{i_{j}})^{(j)}$'s that  what one gets in coordinates is a volume form
over
$\tilde{X}_{w}(x)$ times some multivalued function on $\tilde{X}_{w}(x)$. The
sheaf of continuous
branches of this function is exactly $\cv_{x}$. It also follows from the
construction that the
singularities of this function are concentrated on $
\tilde{X}_{v}(x)\;,v< w$ as well as on the singularities of the sections of
$\co(\cl,\lambda)$. In
other words, the map $\sigma_{j}:\tilde{X}_{w}(x)\rightarrow X_{w}(x)$ makes
$\cv_{x}$ into a local
system over $\dot{X}_{w}(x)\cap F^{0}$ which we  also denote by $\cv_{x}$.

5.Let $H_{l(w)}(\cv_{x})$ be the highest homology group of
$\dot{X}_{w}(x)\cap F^{0}$ with coefficients in $\cv_{x}$. There arises a
vector bundle over
$F^{0}$ with the fiber over $x\in F^{0}$ equal to
$H_{l(w)}(\cv_{x})$. Denote this bundle by $\ch_{l(w)}$. It is equipped
with the canonical Gauss-Manin connection and, therefore, $\gtg$ acts on its
sections. Another
important point is that there is a map $\ch_{l(w)}\rightarrow Vol^{\ast}$,
determined by the
fiberwise integration. It follows that pushing all sheaves in
(\ref{almost_final_2}) forward one
obtains
\[\Gamma(\tilde{F}^{0},\co(\cl,\lambda)\otimes\ch_{l(w)})
\rightarrow
\Gamma(\tilde{F}^{0},\co (\cl , r_{i_{1}}\cdots r_{i_{l}}\cdot\lambda))\]
or, equivalently,

\begin{equation}
\label{final}
T(r_{i_{l}}\cdots r_{i_{1}}):\; \Gamma(\tilde{F}^{0},\co(\cl,\lambda))
\rightarrow
\Gamma(\tilde{F}^{0},\;\co (\cl , r_{i_{1}}\cdots
r_{i_{l}}\cdot\lambda)\otimes\ch_{l(w)}^{\ast}).
\end{equation}

We have arrived at
\begin{theorem}
\label{th_int_rel_arb_elem}
For each reduced decomposition $w=r_{i_{l}}\cdots r_{i_{1}}$
the map $T(r_{i_{l}}\cdots r_{i_{1}})$ in (\ref{final})
 is $\gtg-$linear.
\end{theorem}

6. The $\gtg-$modules appearing in different sides of (\ref{final}) are of
different size: while
$\Gamma(\tilde{F}^{0},\co(\cl,\lambda))$
may be regarded as a space of (multi-valued) functions on the flag
manifold, it is no longer true for
$\Gamma(\tilde{F}^{0},\;\co(\cl,r_{i_{1}}\cdots
r_{i_{l}}\cdot\lambda)\otimes\ch_{l(w)})$
because there is no reason
to think that $\ch_{l(w)}$ is a line bundle. Nevertheless one may look for a
line subbundle of
 the latter bundle such that the space of its sections is closed with respect
to the action
of $\gtg$. These
subbundles are described via the Gauss-Manin connection on $\ch_{l(w)}$. This
connection makes the
fundamental group $\pi_{1}(F^{0})$ act on  fibers $H_{l(w)}(\cv_{x})$. Denote
by
$P(H_{l(w)}(\cv_{x}))$
 the projectivization of $H_{l(w)}(\cv_{x})$. Each $\pi_{1}(F^{0})$-invariant
element $h$ of
$P(H_{l(w)}(\cv_{x}))$ gives rise to a line subbundle of $\ch_{l(w)}$ which we
denote
$\ch_{l(w)}(h)$. Obviously $\gtg-$ acts on the sections of $\ch_{l(w)}(h)$. The
inclusion
$\ch_{l(w)}(h)\subset \ch_{l(w)}$ gives rise to the projection
$\ch_{l(w)}^{\ast}\subset\ch_{l(w)}(h)^{\ast}$. We have proved the 1st item of
\begin{lemma}
(i) For each reduced decomposition $w=r_{i_{l}}\cdots r_{i_{1}}$ and a
$\pi_{1}(F^{0})$-invariant element $h$ of
$P(H_{l(w)}(\cv_{x}))$ there is a $\gtg-$linear map
\[\Gamma(\tilde{F}^{0},\co(\cl,\lambda))
\rightarrow
\Gamma(\tilde{F}^{0},\;\co (\cl , r_{i_{1}}\cdots
r_{i_{l}}\cdot\lambda)\otimes\ch_{l(w)}^{\ast}(h));\] (ii) There is at least
one
$\pi_{1}(F^{0})$-invariant element  of
$P(H_{l(w)}(\cv_{x}))$.
\end{lemma}

As to the 2nd item, it  was actually  proved at the beginning  of this section.
The repeated
integration in (\ref{simple_case_intertw}) associates a
$\pi_{1}(F^{0})$-invariant element  of
$P(H_{l(w)}(\cv_{x}))$ to any reduced decomposition $w=r_{i_{l}}\cdots
r_{i_{1}}$. The cycle of
integration in (\ref{simple_case_intertw}) looks like a product of some simple
cycles. As was
explained in
sect.\ref{Construction_of_an_Intertwining_Operator_related_to_a_simple root},
 to each 1-dimensional Schubert variety $X_{r_{i}}(x)$
one canonically associates a cycle with coefficients in a local system on
$X_{r_{i}}(x)$
punctured at 3 points. Denote this cycle by $\sigma_{i}(x)$. Set
$\sigma_{i_{l}}\circ\cdots\circ\sigma_{i_{1}}(x_{0})=    \{(x_{1},\ldots
,x_{l}):\;x_{j}\in\sigma_{i_{j}}(x_{j-1}\}$. Since each $\sigma_{i}(x)$ is an
eigenvector of the
fundamental group of $F^{0}$ (the corresponding homology group is
1-dimensional),
the ``product''
$\sigma_{i_{l}}\circ\cdots\circ\sigma_{i_{1}}(x_{0})$ is also. It is not clear
whether it depends on
the reduced decomposition $w=r_{i_{l}}\cdots r_{i_{1}}$. $\Box$

\section {{\bf Solutions to Knizhnik - Zamolodchikov Equations. The classical
case.}}
\label{Solutions_of_Knizhnik_Zamolodchikov_Equations_The_classical_case}
\subsection{{\bf Trigonometric Form of Knizhnik-Zamolodchikov Equations and
Correlation
Functions.}} \label{tr_f_k_z_eq}

1.We, firstly, prepare notations in order to write down the trigonometric form
of
Knizhnik-Zamolodchikov equations. ( In the exposition we will be following
{}~\cite{fr_resh}.) Let
\[\gtg=\oplus_{\alpha\in\Delta}\gtg_{\alpha}\]
be a root decomposition of a simple finite-dimensional Lie algebra $\gtg$.
 Fix an invariant inner product on $\gtg$ and a basis
$\{h_{i}\in\gth,\;g_{\alpha}\in\gtg_{\alpha}:\;1\leq i\leq
n,\;\alpha\in\Delta\}$ of $\gtg$  so that
$(h_{i},h_{j})=\delta_{i,j},\;(g_{\alpha},g_{\beta})=\delta_{\alpha,-\beta}$.
For each
$\mu\in\gth^{\ast}$ denote by $h_{\mu}$ an element of $\gth$ satisfying (and
uniquely determined) by
the condition $(h_{\mu},h)=\mu (h)$.

 Set
\[r=\frac{1}{2}\sum_{i=1}^{n}h_{i}\otimes
h_{i}+\sum_{\alpha\in\Delta_{+}}g_{\alpha}\otimes
g_{-\alpha}.\]
Being an element of $U(\gtg)\otimes U(\gtg)$ $r$ naturally acts on a tensor
product of
 2 $\gtg-$modules. There are $m^{2}$ different ways to make it act on a tensor
product of $m$
$\gtg-$modules via the following $m^{2}$ embeddings of $U(\gtg)^{\otimes 2}$ in
 $U(\gtg)^{\otimes m}$: each of them is associated to a pair of numbers $1\leq
i,j\leq m$ and sends
\[U(\gtg)^{\otimes 2}\ni \omega\mapsto \omega_{ij}\in U(\gtg)^{\otimes n},\]
so that
\[\mbox{if } \omega =
\sum_{s}a_{s}\otimes b_{s}\mbox{ then }\omega_{ij}=\sum_{s}
\overbrace{\underbrace{1\otimes\cdots 1\otimes a_{s}}_{i}\otimes 1\otimes\cdots
\otimes 1\otimes
b_{s}}^{j}\otimes 1\otimes\cdots\otimes 1.\]

For a pair $1\leq i,j\leq m$
introduce the following function in 2 complex variables with values in
$U(\gtg)^{\otimes m}$:
\[r(z_{i},z_{j})=\frac{r_{ij}z_{i}+r_{ji}z_{j}}{z_{i}-z_{j}}.\]

{\em The trigonometric form of the Knizhnik-Zamolodchikov equations} is the
following system of $m$
differential equations on a function $\Psi$ in $m$ complex variables with
values in the tensor
product $V_{1}\otimes\cdots\otimes V_{m}$
of $m$ $\gtg-$modules:

\begin{eqnarray}
\label{k_z_class}
(k+h^{\vee} )
z_{i}\frac{\partial\Psi}{\partial z_{i}}&=&
\{\sum_{j\neq
i}r_{ij}(z_{i},z_{j})-\frac{1}{2}(\lambda_{1}+\lambda_{m+1}+2\rho)^{(i)}\}\Psi,\\
1\leq &i&\leq m,\nonumber
\end{eqnarray}
where $
h^{\vee}$ is the dual Coxeter number of $\gtg$,
$k\in\nc,\;\lambda_{1},\lambda_{m+1}\in\gth^{\ast}$
are regarded as parameters of the system and, finally, for each $\mu
\in\gth^{\ast}$ $\mu^{(i)}$ stands for the  operator acting on
$V_{1}\otimes\cdots\otimes V_{m}$ as $h_{\mu}$ applied to the $i-$th factor of
$V_{1}\otimes\cdots\otimes V_{m}$. To keep track of the parameters we will be
referring to
(\ref{k_z_class}) as $KZ(\lambda_{m+1},\lambda_{1})$.

Let $V_{1},\ldots
,V_{m}$ be highest (lowest) weight modules with highest (lowest)
 weights $\mu_{1},\ldots ,\mu_{m}$ (resp.) and
correponding weight vectors $v_{\mu_{1}},\ldots ,v_{\mu_{m}}$.
The simplest non-trivial solution to (\ref{k_z_class}) is found in the case
when
it is supposed to take values in the line
spanned by $v_{\mu_{1}}\otimes v_{\mu_{2}}\otimes\cdots\otimes v_{\mu_{m}}$.

 \begin{lemma}
In the above notations the function

\begin{eqnarray}
\Psi =&\prod_{i<j}&(z_{i}-z_{j})^{2(\mu_{i},\mu_{j}) /
(k+h^{\vee})}(z_{i}z_{j})
^{-(\mu_{i},\mu_{j}) / (k+h^{\vee})}\times\nonumber\\
&\prod_{i}&z_{i}^{(e \lambda_{1}+\lambda_{m+1}+2\rho , \mu_{i}) /
2(k+h^{\vee}}\cdot
v_{\mu_{1}}\otimes v_{\mu_{2}}\otimes\cdots\otimes v_{\mu_{m}}\nonumber
\end{eqnarray}

is a solution to the system (\ref{k_z_class}).
\label {simpl_sol_class}
\end{lemma}

2. Denote by $\hgtg =\gtg\otimes\nc [t,t^{-1}]\oplus\nc c$ the non-twisted
affine Lie algebra
associated
 to $\gtg$. The Knizhnik-Zamolodchikov equations were derived in ~\cite{kn_zam}
 as equations on matrix elements
of certain operators acting between highest weight modules over $\hgtg$. Let us
briefly recall basic
definitions,  referring to the book ~\cite{kac_book} for details.
Fix the inclusion $\gtg\subset\hgtg,\;g\mapsto g\otimes 1$. If
$F_{i},H_{i}, E_{i},\;1\leq i\leq n$ are canonical generators of $\gtg$ then
$F_{i},H_{i}, E_{i},\;0\leq i\leq n$ are canonical generators of $\hgtg$ where
it is set
$E_{0}=g_{-\theta}\otimes t,\;F_{0}=g_{\theta}\otimes t^{-1}\mbox{ ($\theta$
stands for the maximal
root of $\gtg$),}
\; H_{0}=[E_{0},F_{0}].$ This includes
affine Lie algebras into the general theory of Kac-Moody Lie algebras, in
particular produces
the Cartan and 2 Borel subalgebras
$\gth\subset\hgth,\;\gtb^{\pm}\subset\hgtb^{\pm}$,
the
triangular decomposition $\hgtg=\hgtb^{-}\oplus\hgth\oplus\hgtb^{+}$, root
decomposition, the Weyl
group etc., as well as the basic objects the of representation theory: Verma
module, highest weight
modules weight decomposition tec., everything being defined in a natural way.

Let $M(\hat{\lambda})$ ($M(\hat{\lambda})^{c}$)
be a Verma module (contragredient Verma module) over $\gtg$ with highest weight
$\hat{\lambda}\in\hgth^{\ast}$. We will also denote these modules as
$M(\lambda,k)$ (or $M(\lambda,k)^{c}$), where
 $\lambda=\hat{\lambda}\mid_{\gth},\;k=\hat{\lambda}(c)$. The last equality
simply means that the
central element $c$ acts as the multiplication by $k\in\nc$. The number
determined by the last
condition is called the level of representation. Observe also that $\hgtg$ is
naturally graded by
powers of the indeterminate $t$ and this gradation induces the gradation of any
highest weight
module $N$:
\[N=\oplus_{i\in\nz}N[m],\]
where the degree of the highest weight vector is said to be equal to 0.

An affine Lie algebra also possesses modules which are not included into the
general theory of
Kac-Moody Lie algebras. For any $\gtg-$module $V$ set
\[V[z] = V\otimes \nc [z,z^{-1}],\]
\[V((z)) = V\otimes \nc ((z)),\]
where $\nc [z,z^{-1}]$ ($\nc ((z))$ resp.) is a space of Laurent polynomials
(series resp.).
Both are naturally equipped with a $\hgtg-$module structure, the level being 0
in both cases
and graded by powers of $z$.
Note that one can consider the spaces $V[z]z^{\nu},\;V((z)) z^{\nu}$ equipped
with a natural
$\hgtg$-module structure and grading ( $deg(v\otimes z^{n+\nu})=n$ )
 so that the map
\[v\otimes z^{n+\nu}\mapsto v\otimes z^{n}\]
provides the isomorphisms
\[V[z]\approx V[z]z^{\nu},\;V((z))\approx V((z)) z^{\nu},\]
of $\hgtg$-modules.

Consider a ( quadratic ) Casimir operator of $\gtg$
\[\Omega=\sum_{\alpha\in\Delta}g_{\alpha}\otimes g_{-\alpha}
+\sum_{i=1}^{n}h^{i}\otimes h^{i}.\]
If $\Omega$ acts on a $\gtg-$module $V$ as a multiplication by a number, we
denote this number by
$c_{V}$. For example, for a highest weight module with a highest weight $\mu$
this number is equal to
$(\mu+\rho,\mu)$.

3.{\em Vertex operator } $\Phi(z)$ is said to be a $\hgtg-$morphism
\begin{eqnarray}
&\Phi (z)&:\; M(\lambda_{1},k)\rightarrow M^{c}(\lambda_{2},k)\otimes V((z))
z^{\kappa}\label{def_vert_op_1}\\
&\kappa& = \frac{1}{2(k+h^{\vee}}(-(\lambda+\rho,\lambda)+c_{V}),\nonumber
\end{eqnarray}
homogeneous with respect to the above-defined gradation.

We will be often referring to $\Phi (z)$ as  a vertex operator acting from
$M(\lambda_{1})$
to $M(\lambda_{2},k)$.

Ovserve that all $\hgtg-$ as well as  $\gtg-$modules under consideration are
semisimple with respect
to the action of Cartan subalgebra and, therefore, in addition to the
above-defined ``exterior''
gradation possess the ``inner'' gradation: the weight decomposition. Now one
easily classifies
vertex operators.

\begin{lemma}
\[Hom_{\hgtg}(M(\lambda_{1},k);\; M^{c}(\lambda_{2},k)\otimes V((z)))\approx
V^{(\lambda_{1}-\lambda_{2})},\]
where $V^{(\lambda_{1}-\lambda_{2})}$ stands for the weight space related to
the weight
$\lambda_{1}-\lambda_{2}$.
\end{lemma}
{\bf Proof.} The passage to the dual $V^{\ast}$ of $V$ makes the assertion into
the following one:
\[Hom_{\hgtg}(M(\lambda_{1},k)\otimes V^{\ast}[z];\;
M^{c}(\lambda_{2},k))\approx
(V^{\ast})^{(\lambda_{1}-\lambda_{2})}.\]
Set $\hgtb=\hat{\gtnp}\oplus\hat{\gth},\;
\hgtb^{-}=\hat{\gtnm}\oplus\hat{\gth}$.
One proves that
\[ Hom_{\hgtg}(M(\lambda_{1},k)\otimes V^{\ast}[z]\approx
\,Ind_{\hgtb}^{\hgtg}\,
\nc v_{\lambda_{1}}\otimes V^{\ast}[z].\]
Frobenius duality implies
\[Hom_{\hgtg}(M(\lambda_{1},k)\otimes V^{\ast}[z];\;
M^{c}(\lambda_{2},k))\approx
Hom_{\hgtb}(\nc v_{\lambda_{1}}\otimes
V^{\ast}[z];\; M^{c}(\lambda_{2},k)).\]
Dualizing one obtains
\[Hom_{\hgtg}(M(\lambda_{1},k)\otimes V^{\ast}[z];\;
M^{c}(\lambda_{2},k))\approx
Hom_{\hgtb}( (M^{c}(\lambda_{2},k))^{\ast};\;
\nc v_{-\lambda_{1}}\otimes
V[z] ).\]
Since (by the definition)
\[(M^{c}(\lambda_{2},k))^{\ast}\approx Ind_{\hgtb^{-}}^{\hgtg}\,\nc
v_{-\lambda_{2}},\]
one more application of the Frobenius duality implies
\[Hom_{\hgtg}(M(\lambda_{1},k)\otimes V^{\ast}[z];\;
M^{c}(\lambda_{2},k))\approx
Hom_{\hgth}( \nc v_{-\lambda_{2}};\;
\nc v_{-\lambda_{1}}\otimes
V[z] ),\]
completing the proof. $\Box$

For a generic highest weight a Verma module is isomorphic to the corresponding
contragredient
Verma module - this is the case we will be interested in - therefore,
 a composition of vertex operators is
well-defined. Namely, if
\[\Phi_{i}(z_{i}):\;
 M(\lambda_{i},k)\rightarrow M(\lambda_{i+1},k)\otimes V_{i}(z_{i}),\;1\leq
i\leq
m\]
are vertex operators, then there arises an operator
\[\Phi_{m}(z_{m})\circ\cdots\circ\Phi_{1}(z_{1}):\; M(\lambda_{1},k)\rightarrow
M(\lambda_{m+1},k)\otimes (V_{1}\otimes\cdots\otimes V_{m})(z_{1},\ldots
,z_{m}),\]
where
$(V_{1}\otimes\cdots\otimes V_{m})(z_{1},\ldots ,z_{m})$ is said to be the
space of Laurent series in
$z_{1},\ldots ,z_{m}$ twisted by certain powers of $z_{i}$ coming from
(\ref{def_vert_op_1})
 with coefficients in $V_{1}\otimes\cdots\otimes V_{m}$.

The composition $
\Phi_{m}(z_{m})\circ\cdots\circ\Phi_{1}(z_{1})$
 is determined by its matrix elements

\[\langle
v_{m+1}^{\ast},\;\Phi_{m}(z_{m})\circ\cdots\circ\Phi_{1}(z_{1})v_{1}\rangle,
\;
v_{1}\in M(\lambda_{1} ,k),\;v_{m+1}\in M(\lambda_{m+1} ,k)^{\ast}.\]

 Each matrix element is
obviously a formal series with coefficients in $V_{1}\otimes\cdots\otimes
V_{m}$.

{\em A correlation function} is said to be the matrix element of a composition
of vertex operators
related to the vacuum vectors:
\[\langle v_{\lambda_{m+1}}^{\ast},\;
\Phi_{m}(z_{m})\circ\cdots\circ\Phi_{1}(z_{1}) v_{\lambda_{1}}\rangle.\]

Remark that vertex operators can be defined to be acting between not
necessarily Verma, or even
highest weight modules, though in the latter case  existence of a composition
of vertex operators
becomes more subtle.

\begin{theorem}[Knizhnik, Zamolodchikov] (\mbox{see ~\cite{kn_zam}})

 The correlation function
\[\langle v_{\lambda_{m+1}}^{\ast},\;
\Phi_{m}(z_{m})\circ\cdots\circ\Phi_{1}(z_{1})v_{\lambda_{1}}\rangle\]

satisfies the system $KZ(\lambda_{m+1},\lambda_{1})$ ( see (\ref{k_z_class}) ).
\label{corr_fun_sol_class}
\end{theorem}
The proof of Theorem \ref{k_z_class}, which can be found in ~\cite{fr_resh},
implies that the following more
precise assertion is valid.

\begin{lemma} The statement of Theorem \ref{k_z_class} remains valid if
\[\Phi_{m}(z_{m})\circ\cdots\circ\Phi_{1}(z_{1}):\; W\rightarrow
M(\lambda_{m+1},k)\otimes
(V_{1}\otimes\cdots\otimes V_{m})(z_{1},\ldots ,z_{m}),\]
is an intertwiner acting from not necessarily Verma module $W$, provided the
following
conditions on $v_{\lambda_{1}}\in W$
hold
\[H\, v_{\lambda_{1}}=\lambda_{1}(H)\,v_{\lambda_{1}},\]
\[\langle v_{\lambda_{m+1}}^{\ast},\mid
\Phi_{m}(z_{m})\circ\cdots\circ\Phi_{1}(z_{1})\mid
\hat{\gtn}^{+}\,v_{\lambda_{1}}\rangle= 0 .\]
\label{compl_to_th}
\end{lemma}

\subsection{{\bf Main result: Integral Representations of Solutions.}}
\label{Main_result:_Integral_Representations_of_Solutions}
1. As was explained in Introduction it is
plausible that a suitable definition of complex powers of generators of $\hgtg$
as operators acting
on $(V_{1}\otimes\cdots\otimes V_{m})(z_{1},\ldots ,z_{m})$ gives rise to
solutions of KZ equations.
We will give such a definition  provided $V_{i}$'s belong
to a suitable family of
$\gtg-$ modules. We assume that
\[V_{i}=\Gamma (\tilde{F}^{0},\co (\cl_{\nu_{i}},\mu_{i})),\;1\leq i\leq m,\]
where $\mu_{i},\mu_{i}$ are arbitrary complex numbers, module
$\Gamma (\tilde{F}^{0},\co (\cl,\mu))$ was defined
in sect.\ref{Construction_of_Intertwining_Operators}
 and the local system
$\cl_{\nu}$ was defined
in sect.\ref{Construction_of_Intertwining_Operators},
Example\ref{ex_of_loc_syst}.

 Simply
speaking $\Gamma (\tilde{F}^{0},\co (\cl,\mu))$ is a module realized in
multi-valued functions on the
flag manofold. The direct way to this realization is as follows: take
$M(\mu)^{c}$; it is realized in
meromorphic sections of the line bundle $\co(\lambda)$. Having a section of
this bundle fixed one
identifies meromorphic  sections with meromorphic functions
  on the flag manifold, the action of $\gtg$ being given by 1st order
differential operators (see, for example, \cite{feig_fren_0}); then the space
$\Gamma (\tilde{F}^{0},\co(\cl,\mu))$  is
determined by multiplication of meromorphic functions by the multivalued
function
$\phi=\prod_{1\leq j\leq
n}\phi_{j}^{\nu_{j}}$ defined in the above-mentioned
Example\ref{ex_of_loc_syst} the action being
given by the same 1st order differential operators. Therefore an element of
$(V_{1}\otimes\cdots\otimes V_{m})(z_{1},\ldots
,z_{m})$ may (and will ) be regarded as a function of a point on a cartesian
product of $m$ copies
of a flag manifold and complex coordinates $z_{1},\ldots,z_{m}$ .  The
dependence on the latter has
been so far formal but it will in reality turn out to determine a multi-valued
analytic function.

2. One of the properties of modules $\Gamma (\tilde{F}^{0},\co
(\cl_{\nu},\mu))$ is that a group
element $\exp (tg)\in G,\;g\in\gtg $  may be regarded as an operator sending
elements of
$\Gamma (F^{0},\co (\cl_{\nu},\mu))$ to elements of the same space times,
probably, another
multi-valued function. In order to understand this, first, observe
 that the group $G$ acts on meromorphic sections of a line bundle $\co(\mu)$,
corresponding to an
integral highest weight $\mu$. Then one defines an action of $G$ on complex
powers of such sections
as follows: if $A\in G$, $s$ is a section of $\co(\mu)$ and $A\cdot s=f_{A}s$,
where $f_{A}$ is a
function on the flag manifold, then one sets
\[A\cdot s^{\alpha}=( f_{A})^{\alpha}\cdot s^{\alpha}.\]
Similar formula applies to the product $\prod_{i}s_{i}^{\alpha_{i}}$ and since
the latter expressions
generate the space of sections of $\co(\mu)$ for an arbitrary $\mu$ (see the
discussion of complex
powers of line bundles over a flag manifold in
sect.\ref{FlagManifoldsLineBundles})
this completes
the definition of the operator $\exp (tg),\;g\in\gtg$.

Factor-wise application makes $G$ act on the tensor product
$V_{1}\otimes\cdots\otimes V_{m}$.

3.We are now in a position to define an action of complex powers of elements of
$\hgtg$ on some
elements
of $(V_{1}\otimes\cdots\otimes V_{m})(z_{1},\ldots ,z_{m})$. First of all, the
action of $\hgtg$ on
$(V_{1}\otimes\cdots\otimes V_{m})(z_{1},\ldots ,z_{m})$ is defined by means of
a certain Lie
algebra homomorphism $\hgtg\rightarrow U(\gtg)^{\otimes m}$. For example, on
generators of
$\hat{\gtn}^{+}$ it acts as follows:

\begin{eqnarray}
F_{i}&\mapsto& \sum_{j=1}^{m}\underbrace{1\otimes\cdots\otimes 1\otimes
F_{i}}_{j}\otimes
1\otimes\cdots\otimes 1,\;1\leq i\leq n,\label{action_on_corr_fun_1}\\
F_{0}&\mapsto& \sum_{j=1}^{m}z_{j}^{-1}\underbrace{1\otimes\cdots\otimes
1\otimes
g_{\theta}}_{j}\otimes 1\otimes\cdots\otimes 1.\label{action_on_corr_fun_2}
\end{eqnarray}

These formulas combined with the above discussion imply that group elements
$\exp
(t_{i}F_{i}),\;0\leq i\leq n$ act on those elements of
$(V_{1}\otimes\cdots\otimes V_{m})(z_{1},\ldots ,z_{m})$ which are analytic
(probably multi-valued)
functions of $z_{1},\ldots,z_{m}$. Denote the subspace of
 $(V_{1}\otimes\cdots\otimes V_{m})(z_{1},\ldots ,z_{m})$ consisting of
analytic
(multi-valued )functions of $z_{1},\ldots,z_{m}$ by
$(V_{1}\otimes\cdots\otimes V_{m})(z_{1},\ldots ,z_{m})^{fun}$.

{\bf Definition.} For $\psi\in (V_{1}\otimes\cdots\otimes V_{m})(z_{1},\ldots
,z_{m})^{fun}$
,$A\in End ((V_{1}\otimes\cdots\otimes V_{m})(z_{1},\ldots ,z_{m})^{fun})$,
$k\in\nc$
 set
\begin{equation}
\label{def_compl_pow}
A^{\beta}\cdot\psi = \Gamma (-\beta)^{-1}\int\{\exp (-tA)\psi \}t^{-\beta-1}dt,
\end{equation}
where the contour is an arbitrary element of the local system determined by
single-valued branches
of the function $\{\exp (-tA)\psi \}$ if the latter expression makes sense.

The above discussion shows that the definition makes sense at least for
elements of $\hgtg$ while
the following lemma shows that the definition implies several natural
properties.

\begin{lemma} The following relations hold provided both sides of them make
sense.

\[\mbox{(i) }A^{\beta}=A^{\beta-n}\cdot A^{n},\;k\in\nc,n\in\nn.\]

(ii) (the binomial theorem)
 if $[A,B]=0$, then
\[(A+B)^{\beta}=\sum_{j=0}^{\infty}\frac{\beta (\beta-1)\cdots(\beta-j+1)}{j!}
 A^{\beta -j}B^{j};\]

\[\mbox{(iii) }[H_{i},F_{j}^{\beta}]=-\beta a_{ij}F_{j}^{\beta},\;k\in\nc;\]

\[\mbox{(iv) }[E_{i},F_{j}^{\beta}]= \delta_{i,j}\beta
F_{j}^{\beta-1}(H_{j}-\beta+1),\;\beta\in\nc\]
\label{form_prop_class}
\end{lemma}
{\bf Proof} is a matter of routine calculations.

{\bf 4. The main result.}
Let $\Psi\in (V_{1}\otimes\cdots\otimes V_{m})(z_{1},\ldots ,z_{m})^{fun}$ be a
correlation function
coming from the composition of vertex operators
\[\Phi_{m}\circ\cdots\circ\Phi_{1}:\; M(\lambda_{1},k)\rightarrow
M(\lambda_{m+1},k)\otimes
(V_{1}\otimes\cdots\otimes V_{m})(z_{1},\ldots ,z_{m}).\]
Let
$w=r_{i_{l}}\cdots r_{i_{2}}r_{i_{1}}\in W$ be a reduced decomposition. Set

\[\beta_{j}=\frac{2(r_{i_{j-1}}\cdots
r_{i_{1}}\cdot\lambda_{1},\alpha_{i_{j}})}
{(\alpha_{i_{j}},\alpha_{i_{j}})}.\]

Set
\[K_{\Psi,w}(t_{1},t_{2},\ldots , t_{l})=\prod_{j=1}^{l}\Gamma
 (-\beta_{j})^{-1}\times\{\exp (-t_{1}F_{i_{1}})\cdots\exp
(-t_{l}F_{i_{l}})\prod_{j=1}^{l}t_{j}
^{-\beta_{j}-1}\Psi\}.\]
Denote by $\cm$ the local system ( over the domain of $
K_{\Psi,w}(t_{1},t_{2},\ldots , t_{l})$ ) of single-valued branches of
$K_{\Psi,w}(t_{1},t_{2},\ldots ,
t_{l})$. The function $
K_{\Psi,w}(t_{1},t_{2},\ldots , t_{l})$ depends on a point
on the cartesian product of flag manifolds as well as on $(z_{1},\ldots
,z_{m})$ as on parameters.
Therefore also does $\cm$. It follows that for every cycle
$\sigma\in H_{l}(\mbox{domain of $K_{\Psi,w}(t_{1},t_{2},\ldots ,
t_{l})$},\;\cm)$ the integral
\[\int_{\sigma}K_{\Psi,w}(t_{1},t_{2},\ldots , t_{l})\,
dt_{1}\, dt_{2}\ldots  dt_{l}\]
is a function of
a point
on the cartesian product of flag manifolds and of $(z_{1},\ldots ,z_{m})$.

\begin{theorem}
For any cycle $\sigma\in H_{l}(\mbox{domain of $K_{\Psi,w}(t_{1},t_{2},\ldots ,
t_{l})$},\;\cm)$ the
function
\begin{equation}
\label{main_res_form}
\int_{\sigma}K_{\Psi,w}(t_{1},t_{2},\ldots , t_{l})\,
dt_{1}\, dt_{2}\ldots  dt_{l}
\end{equation}
 satisfies the system $ KZ(\lambda_{m+1},w\cdot\lambda_{1})$  ( see
(\ref{k_z_class}) ).
\label {main_res_int_class }
\end{theorem}
{\bf Proof} is a formal calculation based on Lemma \ref{compl_to_th} and Lemma
\ref{form_prop_class}
(iii) and (iv). Firstly observe that the composition of vertex operators
$\Phi_{m}\circ\cdots\circ\Phi_{1}$ is naturally extended to a $\hgtg-$morphism:
\begin{eqnarray}
\Phi_{m}\circ\cdots\circ\Phi_{1}:\; &M&(\lambda_{1},k)\hat{\otimes}\nc
((t_{1},\ldots ,t_{l}))
\rightarrow\nonumber\\ &M&(\lambda_{m},k)
\hat{\otimes}\nc ((t_{1},\ldots ,t_{l}))\otimes
(V_{1}\otimes\cdots\otimes V_{m})(z_{1},\ldots ,z_{m}).\nonumber
\end{eqnarray}
Subsequent integration makes it into a $\hgtg-$morphism, denoted symbolically
by
$\int\circ\Phi_{m}\circ\cdots\circ\Phi_{1}$:
\[\int\circ\Phi_{m}\circ\cdots\circ\Phi_{1}:
\; M(\lambda_{1},k)\hat{\otimes}\nc ((t_{1},\ldots ,t_{l}))
\rightarrow M(\lambda_{m},k)\otimes
(V_{1}\otimes\cdots\otimes V_{m})(z_{1},\ldots ,z_{m}).\]
The function
\[\int_{\sigma}K_{\Psi,w}(t_{1},t_{2},\ldots , t_{l})\,
dt_{1}\, dt_{2}\ldots  dt_{l}\]
is equal to the matrix element
\[\langle v_{\lambda_{m+1}},\; \int\circ\Phi_{m}\circ\cdots\circ\Phi_{1}
\{\exp (-t_{l}F_{i_{l}})\cdots\exp (-t_{1}F_{i_{1}})\prod_{j=1}^{l}t_{j}
^{-\beta_{j}-1}
v_{\lambda_{1}}\rangle\]
 of the morphism $\int\circ\Phi_{m}\circ\cdots\circ\Phi_{1}$ because of
the tautological equality
\begin{eqnarray}&\int_{\sigma}&K_{\Psi,w}(t_{1},t_{2},\ldots , t_{l})\,
dt_{1}\, dt_{2}\ldots  dt_{l}=\nonumber\\
&\int_{\sigma}&\langle v_{\lambda_{m+1}}^{\ast},\;
\Phi_{m}(z_{m})\circ\cdots\circ\Phi_{1}(z_{1})
\{\exp (-t_{l}F_{i_{l}})\cdots\exp (-t_{1}F_{i_{1}})\prod_{j=1}^{l}t_{j}
^{-\beta_{j}-1}
v_{\lambda_{1}}\rangle dt_{1}\ldots dt_{l}.\nonumber
\end{eqnarray}
Now the choice of the exponents $\beta_{1},\ldots,\beta_{l}$ and Lemma
\ref{form_prop_class}
(iii) and (iv) shows that the assumption of Lemma \ref{compl_to_th} is valid
for the vector
\[\{\exp (-t_{l}F_{i_{l}})\cdots\exp (-t_{1}F_{i_{1}})\prod_{j=1}^{l}t_{j}
^{-\beta_{j}-1}
v_{\lambda_{1}}\in M(\lambda_{1},k)\hat{\otimes}\nc ((t_{1},\ldots ,t_{l}))\]
(c.f. similar reasoning
in Introduction, formulas (\ref{comm_s_v_0},\ref{comm_s_v}) ), which completes
the proof. $\Box$

We point out that this proof imitates the following situation:

the function
\[\int_{\sigma}K_{\Psi,w}(t_{1},t_{2},\ldots , t_{l})\,
dt_{1}\, dt_{2}\ldots  dt_{l}\]
is a matrix element of the operator $\Phi_{m}\circ\cdots\circ\Phi_{1}$ related
to a singular vector
$F_{i_{l}}^{\beta_{l}}\cdots F_{i_{1}}^{\beta_{1}}\, v_{\lambda_{1}}$.

\subsection {{\bf Explicit Formulas for $\widehat{\gtsl}_{2}$.}}

\label{ExplicitFormulasforgtsl2}
We set $\hgtg=\widehat{\gtsl}_{2}$ and suppose that modules $V_{i}$ are of the
type $V(\mu,\lambda)$
 defined in the Introduction by (\ref{fam_sl2_mod}).
 Under these assumptions one writes out
the integral in (\ref{main_res_form}) in an explicit form.

We denote $E_{1},\;H_{1}\; F_{1}$ the Cartan generators of $\gtsl_{2}$ and set
$E_{0}=F_{1}\otimes
t,\;F_{0}=E_{1}\otimes t^{-1}$ so that $E_{i}\;(F_{i})\;0\leq i\leq 1$ are
generators of maximal
nilpotent subalgebras of $\widehat{\gtsl}_{2}$.
 It is a matter of direct calculation to show that under the identification
\[f(x)\,dx^{-\lambda/2}\approx f(x)\]
 one has
\begin{equation}
\exp (-tE_{1})\, f(x)=f(x-t),\;\;\exp (-tF_{1})\,
f(x)=(-1-tx)^{\lambda}f(\frac{x}{xt+1}).
\label{act_exp_sl2}
\end{equation}

An element of $V(\mu_{1},\lambda_{1})\otimes\cdots\otimes
V(\mu_{m},\lambda_{m})$ is identified with
a function of $(x_{1},\ldots ,x_{m})$
(having the prescribed branching coefficients at coordinate hyperplanes) and a
correlation function $\Psi$ takes the form $\Psi=\Psi(x_{1},\ldots
,x_{m};z_{1},\ldots ,z_{m})$.
Now formulas (\ref{action_on_corr_fun_1}, \ref{action_on_corr_fun_2},
\ref{act_exp_sl2}) imply
\begin{eqnarray}
&\exp (-tF_{1})& \Psi(x_{1},\ldots ,x_{m};z_{1},\ldots ,z_{m})=\nonumber\\
&\{\prod_{i=1}^{m}&(-1-tx_{i})^{\lambda_{i}}\}
\Psi(\frac{x_{1}}{x_{1}t+1},\ldots ,\frac{x_{m}}{x_{m}t+1};z_{1},\ldots ,z_{m})
\label{action_on_corr_fun_sl2_1}\\
&\exp (-tF_{0})& \Psi(x_{1},\ldots ,x_{m};z_{1},\ldots ,z_{m})=\nonumber\\
& &\Psi(x_{1}-tz_{1}^{-1},\ldots ,x_{m}-tz_{m}^{-1};z_{1},\ldots ,z_{m})
\label{action_on_corr_fun_sl2_2}
\end{eqnarray}
These relations can be easily iterated in order to obtain formulas for, for
example
\[\cdots\exp (-t_{2}F_{0})\exp (-t_{1}F_{1})\,\Psi(x_{1},\ldots
,x_{m};z_{1},\ldots ,z_{m}).\]
It is
especially so in the case when $\Psi$ is set to be equal to the ``simplest''
correlation function provided by Lemma \ref{simpl_sol_class}, which is,
actually, a monomial as a
function of $x'$s.
 Now observe that it is all we need, since the Weyl group of
$\widehat{\gtsl}_{2}$ is a free group
generated by 2 reflections $r_{0},\; r_{1}$ and, therefore, each element is
uniquely expanded into
either
\[\cdots r_{0}r_{1}\]
or
\[\cdots r_{1}r_{0}.\]
Let us now write down the results of direct calculations. Set
\[w=\underbrace{r_{\epsilon}\cdots r_{0}r_{1}}_{l},\]
where $\epsilon \equiv l(mod\,2)$;
\begin{eqnarray}
\Psi^{(0)} =&\prod_{i<j}&(z_{i}-z_{j})^{\mu_{i}\mu_{j} / (k+2)}(z_{i}z_{j})
^{\mu_{i}\mu_{j} / 2(k+2)}\times\nonumber\\
&\prod_{i}&z_{i}^{( \lambda_{1}+\lambda_{m+1}+2) \mu_{i} /
2(k+2)};\nonumber
\end{eqnarray}
\[\Psi^{(l)}=\exp (-t_{l}F_{\epsilon})\exp (-t_{l-1}F_{1})\cdots
\exp (-t_{2}F_{0})\exp (-t_{1}F_{1}) \Psi^{(0)}.\]
Observe that $\Psi^{(0)}$ is the solution provided by Lemma
\ref{simpl_sol_class} in the case when
$v_{\mu_{i}},\;1\leq i\leq m$ are highest weight vectors ( of the weight
$\mu_{i},\;\mu_{i}\in\nc$ )
 and
 \[\lambda_{1}=\lambda_{m+1}+\mu_{1}+\cdots\mu_{m}.\]
(Recall that $k$ is a level and a dual Coxeter number for $\gtsl_{2}$ is equal
to 2.)

Formulas (\ref{action_on_corr_fun_sl2_1}, \ref{action_on_corr_fun_sl2_2}) then
give:
\begin{equation}
\label{corrfunforsl2}
\Psi^{(l)}=\prod_{i=1}^{m}\{P_{i}^{(l)}(x_{i};z_{i};t_{1},\ldots
,t_{l})\}^{\mu_{i}} \Psi^{(0)},
\end{equation}
where
\[P_{i}^{(l)}(x_{i};z_{i};t_{1},\ldots ,t_{m})=
\sum_{j=0}^{l}A_{j}(x_{i};z_{i})\sigma_{j}^{(l)}(t_{1},\ldots ,t_{l}),\]
where finally $A_{j}(x;z)$ is given by
\[A_{4i}(x;z)=-z^{-i},\;A_{4i+1}(x;z)=-z^{-i}x,\;A_{4i+2}(x;z)
=z^{-i-1},\;A_{4i+3}(x;z)=z^{-i-1}x;\]
and
\[\sigma_{j}^{(l)}=\sum_{0\leq i_{1} <i_{2}<\cdots <i_{j}< l/2}
t_{2i_{1}+1}t_{2i_{2}+1}\cdots
t_{2i_{j}+1}.\]

{\bf Example.}
\begin{eqnarray}
&\Psi^{(5)}&=\prod_{i=1}^{m}\{-1-x_{i}(t_{1}+t_{3}+t_{5})+z_{i}^{-1}(t_{1}t_{2}+t_{1}t_{4}+t_{3}t_{4})
+\nonumber\\
&z_{i}^{-1}x_{i}&(t_{1}t_{2}t_{3}+t_{1}t_{2}t_{5}+t_{1}t_{4}t_{5}+t_{3}t_{4}t_{5})
-z_{i}^{-2}t_{1}t_{2}t_{3}t_{4}-z_{i}^{-2}x_{i}t_{1}t_{2}t_{3}t_{4}t_{5}\}^
{\mu_{i}}\times
\nonumber\\
&\prod_{i<j}&(z_{i}-z_{j})^{\mu_{i}\mu_{j} / (k+2)}(z_{i}z_{j})
^{\mu_{i}\mu_{j} / 2(k+2)}\times\nonumber\\
&\prod_{i}&z_{i}^{( \lambda_{1}+\lambda_{m+1}+2) \mu_{i} /
2(k+2)};\label{ex_corr_f_sl2}
\end{eqnarray}

Then in notations of Theorem \ref{main_res_int_class } and
(\ref{corrfunforsl2})
 one gets
\begin{equation}
\label{kernelforsl2}
K_{\Psi^{(0)},w}(t_{1},t_{2},\ldots ,
t_{l})=\Psi^{(l)}\times\prod_{i=1}^{l}t_{i}
^{\lambda_{1}+(l-i)(k+2)+1}.
\end{equation}

The function $K_{\Psi^{(0)},w}(t_{1},t_{2},\ldots , t_{l})$ (as a function of
$t$'s)
is multi-valued and branches at
coordinate hyperplanes $t_{i}=0,\;1\leq i\leq m$ and at
$m$ hypersurfaces of the order $l$ given by
\[P_{i}^{(l)}(x_{i};z_{i};t_{1},\ldots ,t_{l})=0.\]
 Let $\cm$ be the local system of continuous branches of
 $K_{\Psi^{(0)},w}(t_{1},t_{2},\ldots , t_{l})$
over the domain of
$K_{\Psi^{(0)},w}(t_{1},t_{2},\ldots , t_{l})$: $D$ . Then our main result
takes the form:

for any $\sigma\in H_{l}(D,\;\cm)$ the integral
\[\int_{\sigma}K_{\Psi^{(0)},w}(t_{1},t_{2},\ldots , t_{l})\,
dt_{1}\, dt_{2}\ldots  dt_{l}\]
satisfies the system $ KZ(\lambda_{m+1},w\cdot\lambda_{1})$  ( see
(\ref{k_z_class}) ),
$K_{\Psi,w}(t_{1},t_{2},\ldots , t_{l})$ being given by (\ref{corrfunforsl2},
\ref{kernelforsl2}).

It is worth mentioning that starting with the correlation function given by
Lemma
\ref{simpl_sol_class}
  for $m=2$ at the 1st step (
$l=1$ one obtains a Gauss hypergeometric function. Increasing of $m$ or of the
number of
steps gives other remarkable special functions which, besides KZ equations,
satisfy some other
differential equations (see sect.\ref{Complex_Powers_Lie_Special_Func}).

\section {{\bf Solutions to $q-$ Knizhnik - Zamolodchikov Equations. }}

\label{Solutions_of_q_Knizhnik_Zamolodchikov_Equations}
  I.Frenkel and N.Reshetikhin have derived (see \cite{fr_resh})
a quantum analogue of the trigonometric form of
KZ equations. Their approach is based on the observation that the
representation theory underlying the
classical KZ equations is to a large extent analogous  to the representation
theory of affine quantum groups.

\subsection{{\bf Quantum Groups and their Representations}}
 The material of this secton is fairly standard. Usually the reference is the
work ~\cite{dc_kac}.

 1. We first recall the definition of the Jimbo-Drinfeld quantum group.
For $q\in\nc,\;d\in\nz$ set:
\[[n]_{d}=\frac{1-q^{2nd}}{1-q^{2d}},\]
\[[n]_{d}!=[n]_{d}\cdots [1]_{d},\]
\[\binq = \frac{[n]_{d}\cdots [n-j+1]_{d}}{[j]_{d}!},\]
 omitting the subscript if $d=1$.
  As usual, $A=(a_{ij}),\,1\leq i,j\leq n$ stands for a generalized
symmetrizable
Cartan matrix ,
symmetrized by non-zero integers $d_{1},\ldots, d_{n}$
 such that $d_{i}a_{ij}=d_{j}a_{ji}$ for all $i,j$.

Now  if $\gtg$ is a Kac-Moody Lie algebra associated to $A$
{\em the Drinfeld -Jimbo quantum group $U_{q}(\gtg),\;q\in\nc$}
is said to be
a Hopf algebra with antipode $S$, comultiplication $\Delta$
 and 1 on
generators $E_{i}, F_{i}, K_{i}, K_{i}^{-1},\,0\leq i\leq n$ and defining
relations

\begin{equation}
\label{q_1}
K_{i}K_{i}^{-1}=K_{i}^{-1}K_{i}=1,\;\;K_{i}K_{j}=K_{j}K_{i},
\end{equation}
\begin{equation}
\label{q_2}
K_{i}E_{j}K_{i}^{-1}=q^{a_{ij}}_{i}E_{j},\;\;K_{i}F_{j}K_{i}^{-1}=q^{-a_{ij}}_{i}F_{j},\;
q_{i}=q^{d_{i}},
\end{equation}
\begin{equation}
\label{q_3}
E_{i}F_{j}-F_{j}E_{i}=\delta_{ij}\frac{K_{i}-K_{i}^{-1}}{q_{i}-q_{i}}
,\; q_{i}=q^{d_{i}},
\end{equation}
\begin{equation}
\label{q_4}
\sum_{\nu =0}^{1-a_{ij}}(-1)^{\nu}q_{i}^{\nu(\nu - 1+ a_{ij})}
\left[ \begin{array}{c}1-a_{ij}\\\nu \end{array}\right]_{d_{i}}
E^{1-a_{ij}-\nu}E_{j}E_{i}^{\nu} = 0\;\;(i\neq j)
,
\end{equation}

the comultiplication being given by
\begin{equation}
\label{comult}
\Delta E_{i}=E_{i}\otimes 1+K_{i}\otimes E_{i},
\Delta F_{i}=F_{i}\otimes K_{i}^{-1}+1\otimes F_{i},
\Delta K_{i}=K_{i}\otimes K_{i},
\end{equation}
and antipode - by
\begin{equation}
\label{antip}
S E_{i}=-K_{i}^{-1}E_{i},
S F_{i}=-F_{i}K_{i},
S K_{i}=K_{i}^{-1}.
\end{equation}

The relations admit the $\nc-$algebra anti-automorphism $\omega$
\begin{equation}
\label{autom_q}
\omega E_{i}=F_{i},\;\omega F_{i}=E_{i},\;\omega K_{i}=K_{i}
\end{equation}

 Set $U^{+}_{q}(\gtg)\;
(U^{-}_{q}(\gtg))$
 equal to the subalgebra, generated by $E_{i}\;(F_{i}\; resp.)\;(1\leq i\leq
n)$
and $U^{0}_{q}(\gtg)=\nc[K_{1}^{\pm 1},\ldots
,k_{n}^{\pm 1}]$.
One may check  that the multiplication induces an isomorphism of linear spaces
\begin{equation}
\label{triang_decomp_q}
U_{q}(\gtg)\approx U^{-}_{q}(\gtg)\otimes U^{0}_{q}(\gtg)\otimes
U^{+}_{q}(\gtg).
\end{equation}

  2. Let $\lambda\in  \gth^{\ast}$. {\em A Verma module} $M_{q}(\lambda)$ is
said to be a
$U_{q}(\gtg)-$module on one generator $v_{\lambda}$ satisfying the following
conditions

\[U^{+}v_{\lambda}=0,\,K_{i}v_{\lambda}=q_{i}^{(\lambda,\alpha_{i})}v_{\lambda}\,(i=1,\ldots,n),\]

\[M_{q}(\lambda)\mbox{ is a free $U^{-}_{q}(\gtg)-$module on the generator
$v_{\lambda}$}.\]

Such a module exists and unique, which follows almost immediately from
(\ref{triang_decomp_q}).
 We will call $v_{\lambda}$ a highest weight ( or vacuum )
 vector. An example of a highest weight module
is the contragredient Verma module $M_{q}^{c}(\lambda)$ which is defined as in
the classical case:
first define the dual to the Verma module by means of the antipode $S$ (see
(\ref{antip}) )
 and then ``twist'' the action
by the antiinvolution $\omega$ (see (\ref{comult}
) ).
We say that an element $w$ of a $U_{q}(\gtg)-$module is a weight vector of the
weight $\mu\in P$
if it satisfies $K_{i}v=q_{i}^{\mu (H_{i})}$. A Verma module as well as its
submodules and quotients
is graded by finite-dimensional weight spaces.

Any quotient $V$ of $M_{q}(\lambda)$ is called {\em a highest weight module}.
Theory of highest weight modules over quantum groups  for generic $q$ is
parallel to that over
$\gtg$. A couple of examples are:

(i) one establishes a 1-1 correspondence between singular vectors (those
annihilated by
$U_{q}^{+}(\gtg)$) and morphisms of Verma modules;

(ii) one proves that  $M_{q}(\lambda)$ has a unique
maximal proper submodule $J_{q}(\lambda)$ and, as was shown in \cite{lus},  the
irreducible quotient
$L_{q}(\lambda)=M_{q}(\lambda)/J_{q}(\lambda)$ is a deformation of the
irreducible $\gtg-$module
with highest weight $\lambda$.

Another important (though simple) observation is that the easily checked
relation
\begin{equation}
\label {comm_motiv}
[E_{i},F_{j}^{m}]=
\delta_{i,j}\frac{q^{m}_{i}-q^{-m}_{i}}{q_{i}-q^{-1}_{i}}F_{i}^{m-1}
\frac{K_{i}q_{i}^{-m+1}-K_{i}q_{i}^{m-1}}{q_{i}-q_{i}^{-1}}
\end{equation}

implies that provided $\lambda (H_{i})+1 \in\nn$ the vector
$F_{i}^{\lambda(H_{i})+1}
v_{\lambda}$ is singular and, therefore, determines a
morphism of a  certain Verma module in $M_{q}(\lambda).$
 In \cite{mal} it was used to derive a quantum analogue of the singular vector
formula, here it will serve as a motivation for a $q-$ integral representation
of $q-$correlation
functions (See Introduction.)

 3. Recall that the definition of a vertex operator had 2
representation-theoretic
ingredients: the  highest weight modules and modules of the type $V[z]\mbox{ or
}V((z))$.
 The latter do not always admit $q-$ deformation but they always do so
in the case when $\gtg=\gtsl_{n}$.
The corresponding construction is as follows.
 Firstly, note that an alternative way to define the
module $V[z]$ for a $\gtg-$ module $V$ is to consider the 1-parametric family
of $\gtg-$ modules
obtained by
the pull-back $\pi^{\ast}(t)V$
of $V$ with respect to the 1-parametric family ($t\in\nc$) of Lie algebra
homomorphisms
(evaluation maps)
\[\pi(t):\;\hgtg\rightarrow \gtg,\;g\otimes z^{m}\mapsto t^{n}g,\;c\mapsto 0.\]
The above-mentioned theorem of Lusztig (see \cite{lus})
implies that (at least  finite-dimensional) $V$
admits a $q-$deformation $V_{q}$. As was shown, for example in \cite{jim},
if $\gtg=\gtsl_{n}$ then
 the evaluation map $\pi(t): \widehat{\gtsl_{n}}\rightarrow \gtsl_{n}\subset
\gtgl_{n}$
also admits a $q-$deformation to the map
 $\pi^{\ast}(t):U_{q}(\widehat{\gtsl_{n}})\rightarrow U_{q}(\gtgl_{n})$,
 which gives a 1-parametric family of $U_{q}(\widehat{\gtsl_{n}})-$ modules
$\pi^{\ast}_{q}(t)V_{q}$. This
produces the $q-$analogues of $V[z]\mbox{ and }V((z))$:
\[V_{q}[z]=\pi^{\ast}_{q}(t)V_{q}\otimes \nc [z,z^{-1}]
,\; V_{q}((z))=\pi^{\ast}_{q}(t)V_{q}\otimes \nc ((z,z^{-1})).\]

In general the evaluation map cannot be deformed. In what follows one should
think either that
$\gtg=\gtsl_{n}$ or that the module $V[z]$ admits a $q-$deformation.

\subsection {{\bf $q-$Vertex Operators and $q-$Correlation Functions}}
We proceed in complete accordance with the classical case.

{\em Vertex operator } is said to be a $U_{q}(\hgtg)-$ homomorphism
 (exactly as in (\ref{def_vert_op_1}) )
\[\Phi (z):\;M_{q}(\lambda_{1},k)\rightarrow M^{c}_{q}(\lambda_{2},k)\otimes
V_{q}
\otimes \nc((z,z^{-1})) z^{\kappa} .\]

One also proves that
\begin{equation}
Hom_{U_{q}(\hgtg)}(M_{q}(\lambda_{1},k), M^{c}_{q}(\lambda_{2}, k)\otimes
V_{q}((z))\,)\approx
V_{q}^{(\lambda_{1}-\lambda_{2})}
\label{class_quant_vert_op}
\end{equation}

In the same way one defines a composition of vertex operators
\begin{equation}
\label{compvertoperdef}
\Phi_{i}(z_{i}):\; M_{q}(\lambda_{i},k)\rightarrow M_{q}(\lambda_{i+1},k)
\otimes ( V_{q})_{i}(z_{i}),\;1\leq i\leq m
\end{equation}
as an operator
\[\Phi_{m}(z_{m})\circ\cdots\circ\Phi_{1}(z_{1}):\;
M_{q}(\lambda_{1},k)\rightarrow
M_{q}(\lambda_{m},k)\otimes ((V_{q})_{1}\otimes\cdots\otimes
(V_{q})_{m})(z_{1},\ldots ,z_{m}),\]
where
$((V_{q})_{1}\otimes\cdots\otimes (V_{q})_{m})(z_{1},\ldots ,z_{m})$
is said to be the space of Laurent series in
$z_{1},\ldots ,z_{m}$ with coefficients in $((V_{q})_{1}\otimes\cdots\otimes
(V_{q})_{m})$. It is
determined by its matrix elements

\[\langle
v_{m+1}^{\ast},\;\Phi_{m}(z_{m}\circ\cdots\circ\Phi_{1}(z_{1})v_{1}\rangle,\]

where $v_{1}\in M_{q}(\lambda_{1} ,k),\;v_{m+1}\in M_{q}(\lambda_{m+1}
,k)^{\ast}$.
The matrix element related to vacuum vectors
$v_{1}=v_{\lambda_{1}},\;v_{m+1}=v_{\lambda_{m+1}}$
is said to be a {\em $q-$correlation function}.

I.Frenkel and N.Reshetikhin have proved that a $q-$correlation function
satisfies a certain system
 of $q-$difference equations.
Let $\crr$ be the universal $\crr-$matrix of $U_{q}(\hgtg)$.
 Denote by $R_{V,W}(z)$ its image as an operator acting from $V[z]\otimes W[1]$
to
$V((z))\otimes W[1]$.

\begin{theorem}
\label {k_zam_quant}
Set $p=q^{2(k+h^{\vee})}$.
The $q-$correlation function
\[\Psi(z_{1},\ldots ,z_{m}):
\langle
v_{m+1}^{\ast},\;\Phi_{m}(z_{m})\circ\cdots\circ\Phi_{1}(z_{1})v_{1}\rangle\]
satisfies the following system of $q-$difference  equations ($1\leq j\leq m$ ):

\begin{eqnarray}
&&\Psi(z_{1},\ldots,z_{j-1},pz_{j},z_{j+1},\ldots,z_{m})=
R_{j-1,j}(z_{j-1}/pz_{j})^{-1}\circ\cdots\circ
R_{1,j}(z_{1}/pz_{j})^{-1}\times\nonumber \\
& &q^{-(\lambda_{1}+\lambda_{m+1}+2\rho)_{j}}\times
R_{j,m}(z_{j}/z_{m})\circ\cdots\circ
 R_{j,j+1}(z_{j}/z_{j+1})
\Psi(z_{1},\ldots,z_{m}),\label{explformqkzeqrm}
\end{eqnarray}
where $R_{ij}(z)$ stands for $R_{(V_{q})_{i}, (V_{q})_{j}}(z)$.
\end{theorem}

One observes that it is not necessary to consider vertex operators only acting
between highest
weight modules and
as in the classical case it follows from the proof of I.Frenkel and
N.Reshetikhin that the following
more precise assertion is valid.

\begin{lemma} The statement of Theorem \ref{k_zam_quant} remains valid if
\[\Phi_{m}(z_{m})\circ\cdots\circ\Phi_{1}(z_{1}):\; W\rightarrow
M_{q}(\lambda_{m+1},k)\otimes
(V_{1}\otimes\cdots\otimes V_{m})(z_{1},\ldots ,z_{m}),\]
is an intertwiner acting from not necessarily Verma module $W$, provided the
following
conditions on $v_{\lambda_{1}}\in W$
hold
\[K_{i}\, v_{\lambda_{1}}=q_{i}^{\lambda_{1}(H_{i})}\,v_{\lambda_{1}},\]
\[\langle v_{\lambda_{m+1}}^{\ast},\mid
\Phi_{m}(z_{m})\circ\cdots\circ\Phi_{1}(z_{1})\,
U_{q}^{+}(\hgtg)\,v_{\lambda_{1}}\rangle=0 .\]
\label{compl_to_th_quant}
\end{lemma}

In the sequel we will be considering the $q$KZ system (\ref{explformqkzeqrm})
in the {\em normalized form} meaning that $R-$
matrix is normalized so that
\begin{equation}
R_{ij}(z)\,v_{i}\otimes v_{j}= v_{i}\otimes v_{j}.
\label{norm_cond}
\end{equation}
Remark that this transformation's only impact is that a correlation function
has to be multiplied
by a certain universal factor in order to produce a solution to
(\ref{explformqkzeqrm},\ref{norm_cond}). ( We do not want to calculate this
factor here.) To keep
track of the parameters we will be referring to
(\ref{explformqkzeqrm},\ref{norm_cond}) as
$qKZ(\lambda_{m+1},\lambda_{1})$.

Again in complete accordance with the classical  case, one deduces from
(\ref{class_quant_vert_op})
that
if $\sum_{1\leq i\leq m}\mu_{i}=\lambda_{1}-\lambda_{2}$
then there is only one ( up to proportionality) composition of vertex operators
 (\ref{compvertoperdef}) and the corresponding correlation function satisfies
\[ \Psi (z_{1},\ldots ,z_{m})=\psi (z_{1},\ldots ,z_{m})\cdot
v_{1}\otimes\cdots\otimes v_{m},\]
for some $\nc-$valued function $\psi (z_{1},\ldots ,z_{m})$,
where $v_{i}$ stands for a highest weight vector of $(V_{q})_{i}$. One easily
shows that the soution to $qKZ(\lambda_{m+1},\lambda_{1})$ produced by this
composition of vertex operators is the following simple
power function
\begin{equation}
\prod_{j=1}^{m}z_{j}^{-(\lambda_{1}+\lambda_{m+1}+2\rho)_{j}/2(k+h^{\vee})}
v_{1}\otimes\cdots\otimes v_{m}.
\label{simpl_sol_quant}
\end{equation}

\subsection{{\bf $q-$Integral Representation of Solutions}}

1. Here we show that at least formally solutions of $q$KZ equations may be
written out in a way
analogous to that in the classical case with all the ingredients of our
integral representations
changed for their $q-$analogues, in particular with the integrals changed for
Jackson integrals. We
are unable to verify the convergence of the obtained Jackson integrals because
we do not have enough
many realizations of quantum groups by difference operators. However we will
derive the necessary
formulas for
$U_{q}(\widehat{\gtsl_{2}})$ and carry out explicit calculations
of $q-$correlation functions analogous to those for
$\widehat{\gtsl_{2}}$.

2. {\em The $q-$ multinomial theorem.} Denote by $\nc^{s}[x_{1},\ldots,x_{m}]$
the algebra of skew
 polynomials. Recall that the latter is said to by a $\nc-$algebra on $m$
generators
$x_{1},\ldots,x_{m}$ and relations $x_{j}x_{i}=q^{2}x_{i}x_{j}$ if $j> i$. As
is well known, the following equality holds in $\nc^{s}[x_{1},\ldots,x_{m}]$:
\begin{equation}
\label{q_comm_vers_mult_th}
(x_{1}+\cdots + x_{m})^{n}=\sum_{i_{1}+\cdots+i_{m}=n}
\frac{[n]!}{[i_{1}]!\cdots [i_{m}]!}x_{1}^{i_{1}}x_{2}^{i_{2}}\cdots
x_{m}^{i_{m}}\; n\in\nn
\end{equation}

3. {\em  Some $q-$calculus.} Suppose $q\in\nc,\;|q|<1$.

3.1 The $q-$ analogue of the usual integral is given by the Jackson integral
related to a finite or semi-infinite ``$q-$interval'',
defined by either
\[\int_{0}^{c}\,f(t)\,d_{q}t=c(1-q^{2})\sum_{n=0}^{\infty}f(cq^{2n})q^{2n}\,
,\;c\in\nc
\]
or
\[\int_{0}^{c\infty}\,f(t)\,d_{q}t=c(1-q^{2})\sum_{n=-\infty}^{\infty}f(cq^{2n})q^{2n}\, ,\;c\in\nc
\]
the usage of $q^{2}$ being, of course, conventional (see \cite{gasp}).
The Jackson integral enjoys some of
the elementary properties of the usual one. Here we point out some of them.

{\em Change of variables.}
 The following is evident
\begin{equation}
\label{ch_var_quant}
\int_{0}^{c\epsilon}\,f(at)\,d_{q}t=a^{-1}\int_{0}^{ac\epsilon}\,f(t)\,d_{q}t,
\end{equation}
where $\epsilon$ is either 1 or $\infty$.

{\em Integration by parts.} Introduce 2 natural $q-$analogues of the
derivative:
\[\dq f(x)=\frac{f(x)-f(q^{2}x)}{x(1-q^{2})},\]
\[\tdq f(x)=\frac{f(x)-f(q^{-2}x)}{x(1-q^{2})}.\]
One immediately proves the
following ``$q-$integration by parts formula'':
\begin{equation}
\label{int_part_quant}
\int_{0}^{c}\,f(x) \tdq g(x)\,d_{q}t=\int_{0}^{c}\,\dq f(x)g(x)\,d_{q}t -
f(c)g(q^{-2}c).
\end{equation}

One also:

easily finds an analogue of (\ref{int_part_quant}) for a semi-infinte interval;

defines a Jackson integral over an interval $[a,c]$;

defines a  Jackson integral over a ``multi-dimesional region''
\[\int\, f(t_{1},\ldots t_{m})\,d_{q}t_{1}d_{q}t_{2}\cdots d_{q}t_{m}\]
as a repeated Jackson integral and proves that if the function $f(t_{1},\ldots
t_{m})$ is naturally
restricted then the result is independent of the order of integration.
 The repeated integration as well as an integral over $[a,c]$
also admits natural analogues of
(\ref{ch_var_quant} , \ref{int_part_quant} ).

In what follows the Jackson integrals will be taken over certain
``$q-$cycles''. We are not much
concerned about the choice of the region of integration here and restrict
ourselves to the
following comment. We will usually be dealing with integrals of the form
\[\int_{0}^{c}\,f(t)t^{\alpha}\,d_{q}t,\]
and it will be important that in the integration by parts formula
(\ref{int_part_quant}) the boundary
term vanishes. Therefore, given the integral
\[\int_{0}^{c}\,f(t)t^{\alpha}\,d_{q}t,\]
the interval $[0,c]$ is said to be {\em a  $q-$cycle} if $f(c)=0$. One produces
analogous definitions
for intervals of other types.

3.2 The following notations are widely used:
\[(a)_{i}=\prod_{j=0}^{i-1}(1-aq^{2j}),\]
\[(a)_{\infty}=\prod_{j=0}^{\infty}(1-aq^{2j}).\]
{\em The commutative version of the $q-$binomial theorem} is the following
identity ( see
\cite{gasp})
 \begin{equation}
\label{comm_vers_q_bin_th}
\sum_{i=0}^{\infty}\frac{(a)_{i}}{(q^{2})_{i}}\,x^{i}=\frac{(ax)_{\infty}}{(x)_{\infty}}.
\end{equation}

4.  Suppose $A$ is an operator acting on some functional space. We are going to
consider expressions of the form
\[\int\,f(tA)\,d_{q}t\; ,\]
which can be understood as operators acting on the same space by
\[\int\,f(tA)\,d_{q}t\,(\Psi)=\int\,f(tA)(\Psi)\,d_{q}t\; ,\]
provided the right hand side of the last equation converges. In what follows it
will be assumed
( or checked ) that
this condition is satisfied and, moreover, this integral possesses all
necessary properties, e.g.
absolute, uniform convergence, integration by parts etc.

Let $\Gamma_{q}(\beta)$ be the usual $q-$gamma function. Recall that it is
almost uniquely determined
by the following functional equation (see \cite{ask})
\begin{equation}
\label{f_eq_p_g_f}
\Gamma_{q}(\beta +1)=[\beta]\Gamma_{q}(\beta).
\end{equation}
Set
\begin{equation}
\label{d_tw_p_g_f}
\tg_{q}(\beta )=q^{-\beta (\beta-1)}\Gamma_{q}(\beta).
\end{equation}
It follows that
\begin{equation}
\label{f_eq_tw_g_f}
\tg_{q}(-\beta +1)=-[\beta]\tg_{q}(-\beta).
\end{equation}
We will also be using the $q-$exponential function
\[\exp_{q}(z)=\sum_{n=0}^{\infty}\frac{z^{n}}{[n]!}.\]

{\bf Definition.} For $\beta\in\nc$ set
\begin{equation}
\label{def_compl_pow_quant}
A^{q,\beta} = \tg_{q} (-\beta)^{-1}\int\,\exp_{q} (-tA)\,t^{-\beta-1}d_{q}t,
\end{equation}
where the integration is carried out over some $q-$cycle.

This definition is a $q-$deformation  of a quite standard  defintion of a
complex power of a linear
operator given by (\ref{def_compl_pow}) and, as the next lemma shows, is
necessary to take into
account the relations within a quantum group. We also wish to point out that
the symbol $A^{q,\beta}$
is  slightly ambiguous: actually $A^{q,\beta}$ stands for a collection of
operators, each of them
being related to the choice of a contour of integration ( see, for example,
(ii) of the next Lemma.)

\begin{lemma} The following relations hold provided both sides of them make
sense.

\[\mbox{(i) }A^{q,\beta}=A^{q,\beta-n}\cdot A^{n},\;k\in\nc,n\in\nn.\]

(ii) (the $q-$multinomial theorem) Fix $m,\;r:1\leq r\leq m$;
 if operators $\{A_{j}\}$ satisfy
 $A_{j}A_{i}=q^{2}A_{i}A_{j},\;1\leq i<j\leq m$, then provided a suitable
choice of the contour
has been made, one has
\[(A_{1}+\cdots +A_{m})^{q,\beta}=\sum_{j=0}^{\infty}\sum_{j_{1}+\cdots +
j_{m-1}=j}
\frac{[\beta][\beta-1]\cdots
[\beta-j+1]}{[j_{1}]!\cdots[j_{m-1}]!}A_{1}^{j_{1}}\cdots
A_{r-1}^{j_{r-1}}A_{r}^{\beta -j}
A_{r+1}^{j_{r}}\cdots A_{m}^{j_{m-1}}\]

\[\mbox{(iii)
}K_{i}F_{j}^{q,\beta}K_{i}^{-1}=q_{i}^{-a_{ij}\beta}F_{j}^{q,\beta};\]

\[\mbox {(iv)}
[E_{i},F_{j}^{q_{j},\beta}]=\delta_{i,j}\frac{q_{j}^{\beta}-q_{j}^{-\beta}}{q_{j}-q_{j}^{-1}}
F_{j}^{q_{j},\beta-1}\frac{K_{j}q^{-\beta+1}_{j}-K_{j}^{-1}q_{j}^{\beta-1}}{q_{j}-q^{-1}_{j}}.\]
\label{form_prop_quant}
\end{lemma}
{\bf Proof} is a matter of straightforward calculations. We scetch it briefly.

(i) follows from the relations:
\[\tdq (t^{-\mu})=[\mu]t^{-\mu-1},\]
\[\dq (\exp_{q} (-tA))=-A \exp_{q} (-tA),\]
\[\tg_{q}(-\mu +1)=-[\mu]\tg_{q}(-\mu),\]
and the integration by parts formula
\[\int\,\exp_{q} (-tA)\,\tdq (t^{-\mu})\,d_{q}t\,=\,\int\,\dq\exp_{q} (-tA)\,
t^{-\mu}\,d_{q}t,\]
( this is the point where the ``$q-$cycle condition'' is used.)

(ii) follows from the case $m=2$ of (\ref{q_comm_vers_mult_th} )
and elementary manipulations with sums
and Jackson integrals.

(iii) and (iv) are proved analogously. For example, (iv) follows from the
term-wise application of
its  integral analogue, which we have already cited several times:
\[[E_{i},F_{j}^{n}]=\delta_{i,j}\frac{q_{j}^{n}-q_{j}^{-n}}{q_{j}-q_{j}^{-1}}
F_{i}^{n-1}\frac{K_{j}q^{-n+1}_{j}-K_{i}^{-1}q_{j}^{n-1}}{q_{j}-q^{-1}_{j}},\;n\in\nn.\]

{\em 5. The main result.} We proceed in a complete accordance with the
analogous section devoted
to classical correlation functions.
Let $\Psi\in (V_{1}\otimes\cdots\otimes V_{m})(z_{1},\ldots ,z_{m})$ be a
correlation function
coming from the composition of vertex operators
\[\Phi_{m}(z_{m})\circ\cdots\circ\Phi_{1}(z_{1}):\;
M_{q}(\lambda_{1},k)\rightarrow
M_{q}(\lambda_{m+1},k)\otimes (V_{1}\otimes\cdots\otimes V_{m})(z_{1},\ldots
,z_{m}).\]
Let
$w=r_{i_{l}}\cdots r_{i_{2}}r_{i_{1}}\in W$ be a reduced decomposition. Set

\[\beta_{j}=\frac{2(r_{i_{j-1}}\cdots
r_{i_{1}}\cdot\lambda_{1},\alpha_{i_{j}})}
{(\alpha_{i_{j}},\alpha_{i_{j}})}.\]

Set
\[K_{\Psi,w}(t_{1},t_{2},\ldots , t_{l})=\prod_{j=1}^{l}\tg_{q}
(-\beta_{j})^{-1}\times\{\exp_{q}(-t_{1}F_{i_{1}})\cdots\exp_{q}(-t_{l}F_{i_{l}})\prod_{j=1}^{l}t_{j}
^{-\beta_{j}-1}\Psi\}.\]

\begin{theorem}
The
integral
\begin{equation}
\label{main_res_form_quant}
\int_{\sigma}K_{\Psi,w}(t_{1},t_{2},\ldots , t_{l})\,
d_{q}t_{1}\, d_{q}t_{2}\ldots  d_{q}t_{l}
\end{equation}
 satisfies (if exists)
the system $qKZ(\lambda_{m+1},w\cdot\lambda_{1})$ ( see
(\ref{explformqkzeqrm},\ref{norm_cond}) ).
\label {main_res_int_quant }
\end{theorem}
{\bf Proof} is a literal repetition of that of Theorem \ref{main_res_int_class
}, where instead of
Lemmas \ref{compl_to_th} and \ref{form_prop_class} (iii), (iv), Lemmas
\ref{compl_to_th_quant} and \ref{form_prop_quant} (iii), (iv) are used. $\Box$

\subsection{{\bf Explicit Formulas for $U_{q}(\widehat{\gtsl_{2}})$ }}
\label{ExplicitFormulasorU_q(widehatgtsl_2}
In this section we carry out an explicit
calculation of the integral (\ref{main_res_form_quant}).

{1. \em $q-$difference operators realization of the algebra
$U_{q}(\gtsl_{2})$.} The algebra
$U_{q}(\gtsl_{2})$ is a Hopf algebra on generators
$E=E_{1},\;K=K_{1},\;F=F_{1}$ and relations
(or
definitions) (\ref{q_1} - \ref{antip}), where $A=(2)$. Direct calculations show
that the following
operators determine a representation of $U_{q}(\gtsl_{2})$ in the space of
functions in 1 variable:
\begin{eqnarray}
E:\,f(x)&\mapsto&
-x\frac{q^{-2\lambda+1}f(q^{2}x)-q^{2\lambda+1}f(x)}{q^{2}-1};
\label{realsl2q_1}\\
K:\,f(x)&\mapsto& q^{-2\lambda}f(q^{2}x) ;
\label{realsl2q_2}\\
E:\,f(x)&\mapsto& \frac{f(x)-f(q^{-2}x)}{(1-q^{-2})x}.
\label{realsl2q_3}
\end{eqnarray}
This action preserves the spaces of polynomials $\nc [x]$, Laurent polynomials
$\nc [x,x^{-1}]$ and
``twisted'' Laurent polynomials $x^{\nu}\nc [x,x^{-1}]$. The last space
produces a (generically
irreducible)  $V_{q}(\nu,\lambda)$- module. The last representation is almost a
deformation of the
 $\gtsl_{2}-$module $V(\nu,\lambda)$
 as well as formulas (\ref{realsl2q_1} - \ref{realsl2q_3}) are almost
deformations
of the formulas (\ref{realsl2} , \ref{act_tens}): they would be actual
deformations if the latter
were composed with the canonical automorphism $E\mapsto F,\;F\mapsto
E,\;H\mapsto -H$.

One also considers the $U_{q}(\widehat{\gtsl_{2}})$-module
$V_{q}(\nu_{1},\lambda_{1})\otimes\cdots\otimes V_{q}(\nu_{m},\lambda_{m})
(z_{1},\ldots ,z_{m})$, the action being determined through the associative
algebra homomorphism
\[U_{q}(\widehat{\gtsl_{2}})\rightarrow U_{q}(\gtsl_{2})^{\otimes m},\]
\[E\mapsto \Delta^{m-1}E,\;F\mapsto \Delta^{m-1}F,\]
\[F_{0}\mapsto E^{(1)}+\cdots +E^{(m)},\]
\[E_{0}\mapsto F^{(1)}+\cdots +F^{(m)},\]
where
\[E^{(i)}=z_{i}^{-1}\,\underbrace{K\otimes\cdots\otimes K}_{i-1}\otimes
E\otimes
\underbrace{1\otimes\cdots\otimes 1}_{m-i},\]
\[F^{(i)}=z_{m-i+1}\,
\underbrace{1\otimes\cdots\otimes 1}_{m-i}\otimes E\otimes
\underbrace{K\otimes\cdots\otimes K}_{i-1}
.\]

{2. \em $q-$ exponent of a $q-$difference operator.}  Recall
that the crucial step in getting an explicit form of the classical integral
representation
(\ref{main_res_form})
 was  evaluation of an exponent of an order 1
differential operator, which is actually a classical problem of the theory of
ordinary differential
equations. We are unaware of a general approach to evaluation of a $q-$
exponent of a $q-$difference
operator which arises in (\ref{main_res_form_quant}).
 However we here report on a straightforward calculation of the quantities
\[\exp_{q}(-tX)\, x_{1}^{\alpha_{1}}x_{2}^{\alpha_{2}}\cdots
x_{m}^{\alpha_{m}},\]
where $X=F$ or $F_{0}$ and $x_{1}^{\alpha_{1}}x_{2}^{\alpha_{2}}\cdots
x_{m}^{\alpha_{m}}$ is regarded
as an element of the module
 $V_{q}(\nu_{1},\lambda_{1})\otimes\cdots\otimes V_{q}(\nu_{m},\lambda_{m})
(z_{1},\ldots ,z_{m})$.

\begin{lemma}
\label{eva_of_q_exp}
(i) If $x^{\alpha}\in V(\lambda,\nu)$ then
\begin{eqnarray}
\exp_{q}(-tF)x^{\alpha}&=&x^{\alpha}\frac{(q^{-2\alpha+2}t/x)_{\infty}}{(q^{2}t/x)_{\infty}}
\label{eva_of_q_exp_1_F},\\
\exp_{q}(-tF_{0})x^{\alpha}&=&x^{\alpha}\frac{(q^{-2\lambda+2\alpha+1}tx/z)_{\infty}}
{(q^{2\lambda+1}tx/z)_{\infty}}
\label{eva_of_q_exp_1_E}.
\end{eqnarray}

(ii)
\begin{equation}
\exp_{q}(-tF)x_{1}^{\alpha_{1}}\cdots x_{m}^{\alpha_{m}}=
x_{1}^{\alpha_{1}}\cdots x_{m}^{\alpha_{m}}
\prod_{i=0}^{m-1}
\frac{(q^{2s_{i}}t/x_{i+1})_{\infty}}{(q^{2r_{i}}t/x_{i+1})_{\infty}},
\label{eva_of_q_exp_m_F}
\end{equation}
where
\[s_{i}=\lambda_{i+2}+\cdots+\lambda_{m}-\alpha_{i+1}-\cdots-\alpha_{m}
+1,\]
\[r_{i}=\lambda_{i+2}+\cdots+\lambda_{m}-\alpha_{i+2}-\cdots-\alpha_{m}
+1,\]
and
\begin{equation}
\exp_{q}(-tF_{0})x_{1}^{\alpha_{1}}\cdots x_{m}^{\alpha_{m}}=
x_{1}^{\alpha_{1}}\cdots x_{m}^{\alpha_{m}}
\prod_{i=0}^{m-1}
\frac{(q^{2\tilde{s}_{i}}tx_{i+1}/z_{i+1})_{\infty}}
{(q^{2\tilde{r}_{i}}tx_{i+1}/z_{i+1})_{\infty}},
\label{eva_of_q_exp_m_E}
\end{equation}
where
\[\tilde{s}_{i}=\alpha_{i+1}+\cdots +\alpha_{1}
-\lambda_{i+1}-\cdots-\lambda_{1}+1/2,\]
\[\tilde{r}_{i}=\alpha_{i}+\cdots+\alpha_{1}+\lambda_{i+1}-\lambda_{i}-\cdots-\lambda_{1}
+1/2.\]
\end{lemma}
{\bf Proof.} All assertions are proved by direct calculations. For example, to
prove
(\ref{eva_of_q_exp_1_F}) one firstly observes, that for $i\in \nn$
\[F^{i}x^{\alpha}=(-1)^{i}q^{2i}\frac{(q^{-2\alpha})_{i}}{(1-q^{2})^{i}}x^{n-i},\]
and then gets
\begin{eqnarray}
\exp_{q}(-tF)x^{\alpha}&=&x^{\alpha}\sum_{i=0}^{\infty}
\frac{(q^{-2\alpha})_{i}}{(q^{2})_{i}}(q^{2}t/x)^{i}\nonumber\\
&=&x^{\alpha}\frac{(q^{-2\lambda+2\alpha+1}t/x)_{\infty}}
{(q^{2\lambda+1}t/x)_{\infty}},
\end{eqnarray}
the last equality following from the commutative version
(\ref{comm_vers_q_bin_th})
of the $q-$binomial theorem.

(\ref{eva_of_q_exp_1_E}) is proved in exactly the same way. As to
(\ref{eva_of_q_exp_m_F}, \ref
{eva_of_q_exp_m_E}), the only difference is that the formula for
$F^{n}x_{1}^{\alpha_{1}}\cdots x_{m}^{\alpha_{m}}$ or
$F_{0}^{n}x_{1}^{\alpha_{1}}\cdots x_{m}^{\alpha_{m}}$ is a little bit more
complicated. For example,
since the elements $E_{(i)}$ satisfy the skew polynomial condition:
\[E^{(j)}
E^{(i)}=q^{2}E^{(i)}E^{(j)}\mbox
{ if }i< j,\]
 $F_{0}^{n}x_{1}^{\alpha_{1}}\cdots x_{m}^{\alpha_{m}}$ can be expanded by
(\ref{q_comm_vers_mult_th}).
 Further application of the commutative version of
the $q-$binomial theorem (\ref{comm_vers_q_bin_th}) as above gives the
required result (\ref{eva_of_q_exp_m_E}). $\Box$

As an immediate consequence of Lemma \ref{eva_of_q_exp} and the definition one
gets
\begin{eqnarray}
F^{q,\mu}x^{\alpha}&=&x^{\alpha-\mu}\frac{q^{-2\mu\alpha}}{\tg_{q}(-\mu)}\int_{0}^{1}\,
t^{-\mu-1}\frac{(q^{2}t)_{\infty}}{(q^{2\alpha+2}t)_{\infty}}\,d_{q}t=\nonumber\\
& &q^{\mu(\mu-2\alpha+1)}\frac{\Gamma_{q}(\alpha+1)}{\Gamma_{q}(\alpha+1-\mu)}
x^{\alpha-\mu},
\label{eva_of_q_comppow_1_F}
\end{eqnarray}

\begin{eqnarray}
F^{q,\mu}x^{\alpha}_{0}&=&x^{\alpha+\mu}z^{-\mu}\frac{q^{-\mu(2\lambda-2\alpha+1)}}{\tg_{q}(-\mu-1)}
\int_{0}^{1}\frac{(q^{2}t)_{\infty}}
{(q^{4\lambda-2\alpha+2}t)_{\infty}}\,d_{q}t=\nonumber\\
&
&q^{2\mu(\lambda-\alpha+1)}\frac{\Gamma_{q}(2\lambda-\alpha+1)}{\Gamma_{q}(2\lambda-\alpha-\mu+1)}
x^{\alpha+\mu}z^{-\mu},\label{eva_of_q_comppow_1_E}
\end{eqnarray}

\begin{equation}
F^{q,\mu}x_{1}^{\alpha_{1}}\cdots x_{m}^{\alpha_{m}}=
x_{1}^{\alpha_{1}}\cdots
x_{m}^{\alpha_{m}}\frac{1}{\tg_{q}(-\mu)}\int\,t^{-\mu-1}
\prod_{i=0}^{m-1}
\frac{(q^{2s_{i}}t/x_{i+1})_{\infty}}{(q^{2r_{i}}t/x_{i+1})_{\infty}}\,d_{q}t,
\label{eva_of_q_compow_m_F}
\end{equation}
\begin{equation}
F_{0}^{q,\mu}x_{1}^{\alpha_{1}}\cdots x_{m}^{\alpha_{m}}=
x_{1}^{\alpha_{1}}\cdots
x_{m}^{\alpha_{m}}\frac{1}{\tg_{q}(-\mu)}\int\,t^{-\mu-1}
\prod_{i=0}^{m-1}
\frac{(q^{2\tilde{s}_{i}}(tx_{i+1})/z_{i+1})_{\infty}}
{(q^{2\tilde{r}_{i}}(tx_{i+1})/z_{i+1})_{\infty}}\,d_{q}t.
\label{eva_of_q_comppow_m_E}
\end{equation}
In integrals (\ref{eva_of_q_compow_m_F}, \ref{eva_of_q_comppow_m_E}) the
integration is supposed to
be carried out over a segment from 0 to a zero of the numerator. One easily
sees that for each of
the integrals there are $m$ essentially
different choices. (It is, of course, closely related to Lemma
\ref{form_prop_quant} (ii).)  An
integral of this type was called by K.Aomoto and K.Mimachi a {\em Jackson
integral of the
Jordan-Pochhammer type}. It is well-known (see, e.g. \cite{gasp}) that if $m=2$
these
integrals produce the
Gauss $q-$hypergeometric function.

3. {\em Integral representations and series expansions for correlation
functions.} In view of
the formula
\ref{simpl_sol_quant} one gets that the function
\[\prod_{j=1}^{m}z_{j}^{-(\lambda_{1}+\lambda_{m+1}+2\rho)_{j}/2(k+h^{\vee})}
x^{\mu_{1}/2}_{1}x_{2}^{\mu_{2}/2}\cdots x^{\mu_{m}/2}_{m}\] is a solution to
(\ref{explformqkzeqrm},\ref{norm_cond})
related to
the composition of vertex operators
\[\Phi_{m}(z_{m})\circ\cdots\circ\Phi_{1}(z_{1}):
M_{q}(\lambda_{1})\rightarrow M_{q}(\lambda_{m+1},k)\otimes ((V(0,\mu_{1})
\otimes\cdots\otimes V(0,\mu_{m}))(z_{1},\ldots ,z_{m})\]

for $\lambda_{1}=\lambda_{m+1}+\mu_{1}+\cdots +\mu_{m}$.
Formulas (\ref{eva_of_q_compow_m_F}, \ref{eva_of_q_comppow_m_E}) combined with
 Theorem \ref{main_res_int_quant } give that the following functions also
satisfy the system
$qKZ(\lambda_{m+1},-\lambda_{1}-2)$ or $qKZ(\lambda_{m+1},2k-\lambda_{1}+2)$
resp.:

\begin{eqnarray}
&
&F^{q,\lambda_{1}+1}\prod_{j=1}^{m}z_{j}^{-(\lambda_{1}+\lambda_{m+1}+2\rho)_{j}/2(k+h^{\vee})}
x^{\mu_{1}/2}_{1}x_{2}^{\mu_{2}/2}\cdots x^{\mu_{m}/2}_{m}=\nonumber\\
& &\prod_{j=1}^{m}z_{j}^{-(\lambda_{1}+\lambda_{m+1}+2\rho)_{j}/2(k+h^{\vee})}
x^{\mu_{1}/2}_{1}x_{2}^{\mu_{2}/2}\cdots
x^{\mu_{m}/2}_{m}\frac{1}{\tg_{q}(-\lambda_{1}-1)}
\times\nonumber\\
& &\int\,t^{-\lambda_{1}-2}
\prod_{i=0}^{m-1}
\frac{(q^{2s_{i}}t/x_{i+1})_{\infty}}{(q^{2r_{i}}t/x_{i+1})_{\infty}}\,d_{q}t,
\label{int_repr _sol_F_quant}\\
&
&F_{0}^{q,k-\lambda_{1}+1}\prod_{j=1}^{m}z_{j}^{-(\lambda_{1}+\lambda_{m+1}+2\rho)_{j}/2(k+h^{\vee})}
x^{\mu_{1}/2}_{1}x_{2}^{\mu_{2}/2}\cdots x^{\mu_{m}/2}_{m}=\nonumber\\
& &\prod_{j=1}^{m}z_{j}^{-(\lambda_{1}+\lambda_{m+1}+2\rho)_{j}/2(k+h^{\vee})}
x^{2\mu_{1}}_{1}x_{2}^{2\mu_{2}}\cdots
x^{2\mu_{m}}_{m}\frac{1}{\tg_{q}(-k+\lambda_{1}-1)}
\times\nonumber\\
& &\int\,t^{-k+\lambda_{1}-2}
\prod_{i=0}^{m-1}
\frac{(q^{2\tilde{s}_{i}}(t(x_{i+1})/z_{i+1})_{\infty}}
{(q^{2\tilde{r}_{i}}(t(x_{i+1})/z_{i+1})_{\infty}}\,d_{q}t,
\label{int_repr _sol_E_quant}
\end{eqnarray}
where
\begin{eqnarray}
s_{i}=\frac{1}{2}(-\mu_{i+1}+\mu_{i+2}
+\cdots \mu_{m})+1,\nonumber\\
r_{i}=\frac{1}{2}(\mu_{i+2}
+\cdots + \mu_{m})+1,\nonumber\\
\tilde{s}_{i}=-\frac{1}{2}(\mu_{i+1}+\cdots \mu_{1}-1),\nonumber\\
\tilde{r}_{i}=-\frac{1}{2}(\mu_{i}+\cdots +\mu_{1}-1)+\mu_{i+1}.\nonumber
\end{eqnarray}

So, we have found solutions to $q$KZ-equations in the form of Jackson integrals
of the
Jordan-Pochhammer type (see formally similar results in \cite{mat}).

In order to get explicitly solutions written as a repeated Jackson integral one
has to evaluate
a result of a repeated application of a $q-$exponent of $F_{1},\;F_{0}$, which
we are unfortunately
unable to do at present. However our method provides a Laurent series expansion
of such solutions.
Really, if $\Psi(z_{1},\ldots ,z_{m})$ is a series expansion of a (not
necessarily
correlation) function then the functions
 \[F^{\mu,q}\Psi(z_{1},\ldots ,z_{m}),\;F_{0}^{\mu,q}\Psi(z_{1},\ldots
,z_{m})\]
 can be evaluated by

(i) expanding the sums
\[F^{\mu,q}=(\sum_{i=1}^{m}F^{(i)})^{\mu,q},\]
\[F^{\mu,q}_{0}=(\sum_{i=1}^{m}E^{(i)})^{\mu,q},\]
using the $q-$binomial theorem (  Lemma \ref{form_prop_quant} (ii) );

(ii) further term-wise application of these expansions to the series
$\Psi(z_{1},\ldots ,z_{m})$
using formulas (\ref{eva_of_q_comppow_1_F}, \ref{eva_of_q_comppow_1_E}).

It is easy to see that starting with $\Psi(z_{1},\ldots ,z_{m})$ polynomial in
$x$'s this process
can be iterated arbitrary number of times, each time giving a converging
series.

\subsection{{\bf Complex Powers of Lie algebra ( Quantum Group ) generators and
Special Functions}}
\label{Complex_Powers_Lie_Special_Func}
1. Let $\gtg$ be a simple finite-dimensional Lie algebra, $G$ be the
corresponding group and
$\Gamma (\tilde{F}^{0},\, \co(\cl_{i},\lambda_{i})),\;1\leq i\leq m$ be
$\gtg-$modules of the type we
were utilizing in
 sect.\ref{Main_result:_Integral_Representations_of_Solutions}. Recall that
the
space $\Gamma (\tilde{F}^{0},\, \co(\cl,\lambda))$ can (and will) be identified
with the space of
multi-valued functions (depending on the local system $\cl$), the action of
$\gtg$ being given by
the same 1st order differential operators
, which produces the realization of the contragredient Verma module with the
highest weight $\lambda$
in (meromorphic) functions on the flag manifold. Also recall that the group
elements $g\in G$ as well
as complex powers Lie algebra elements can be regarded as operators acting from
$\Gamma (\tilde{F}^{0},\, \co(\cl,\lambda))$ to $\Gamma (\tilde{F}^{0},\,
\co(\cl ',\lambda))$
 with some other
$\cl '$ and the same $\lambda$. Since we do not want here to keep track of the
local system, instead
of $\Gamma (\tilde{F}^{0},\, \co(\cl,\lambda))$ we wiil be writing $\cm
(\lambda)$ and slightly
abusing notation we will be saying that for each $g\in G,\; X\in
\gtg,\;\mu\in\nc$ a pair of
operators is defined \[g,\;X^{\alpha}:\;\cm (\lambda)\rightarrow \cm
(\lambda)\]
is defined. (One may also think that $\cm (\lambda)$ is a sum of $\Gamma
(\tilde{F}^{0},\,
\co(\cl,\lambda))$ over ``all $\cl
$''.)

Let $w_{0}$ be the element of the Weyl group of the maximal length and denote
by the same letter its
preimage in $G$. Obviously for any $\lambda$ 1 as a constant function on $F$ is
the vacuum
vector of the module $\otimes_{i=1}^{m}\cm (\lambda_{i})$
 (here the local systems are supposed to be trivial).
It follows that the function $w_{0}1$ is the lowest weight vector ( i.e. it is
annihilated by
$\gtnm$). Therefore $w_{0}1$ is an eigenvector of any Casimir operator of
$\gtg$. It follows that for
any complex number $\alpha$ and any Cartan generator $E_{i}$ the function
$E_{i}^{\alpha}w_{0}1$ is
also. To be definite, let $\Omega_{0},\ldots ,\Omega_{n}$ be a complete set of
Casimir operators
($n+1$ is the rank of $\gtg$ ) and let $\Omega_{i}
w_{0}1=\mu_{i}w_{0}1$. Then one obtains that the function
$E_{i}^{\alpha}w_{0}1$ satisfies the
following system of differential equations:
\begin{equation}
\label{diffeqclassgen}
\Omega_{j}f=\mu_{j}f,\;0\leq j\leq n.
\end{equation}
Observe that the function $E_{i}^{\alpha}w_{0}1$ is homogeneous and therefore
can be identified with
a function of a less number of variables. Under this identification the system
(\ref{diffeqclassgen}) is rewritten in the form explicitly depending on
$\alpha$. Observe that in
all cases, except $\gtg=\gtsl_{2}$, the number of equations in
(\ref{diffeqclassgen}) is less then
the number of variables and one may be interested in finding all other
equations which
$E_{i}^{\alpha}w_{0}1$ satisfies.

2. {\em  Example: $\gtg=\gtsl_{2}$.} In this case $\lambda\in\nc$ and
$\cm(\cl)$ is understood as
$\{f(x)\,dx^{-\lambda/2}\}$ for all $f(x)$ from a suitable family of functions
on $\nc$. The Cartan
generators are represented as follows
\[E=-\frac{d}{dx},\;H=-2x\frac{d}{dx}+\lambda,\;F=x^{2}\frac{d}{dx}-\lambda
x.\]
 The above construction of the lowest weight vector in $\otimes_{i=1}^{m}\cm
(\lambda_{i})$ gives
$x_{1}^{\lambda_{1}}\cdots x_{m}^{\lambda_{m}}$. This is a homogeneous function
of $x_{1},\ldots,x_{m}$ of
homogeneous degree $\lambda_{1}+\cdots +\lambda_{m}$. We identify  a
homogeneous function
$f(x_{1},\ldots,x_{m})$ with the non-homogeneous function of $m-1$ variables
$\tilde{f}(t_{2},\ldots,t_{m}):\;\tilde{f}(t_{2},\ldots,t_{m})=f(1,t_{2},\ldots,t_{m})$. Under this
identification differential operators acting on homogeneous functions of a
fixed degree
are uniquely  transformed into differential operators of less by 1 number of
variables. If for
example, $m=2$ then operators of multiplication $mult(x_{1}),\;mult(x_{2})$ and
differentiation
$\frac{\partial}{\partial x_{1}},\;\frac{\partial}{\partial x_{2}}$ as
operators acting on functions
of degree $\beta$ are transformed as follows:
\[mult(x_{2})\rightarrow mult(t),\;\frac{\partial}{\partial x_{2}}\rightarrow
\frac{d}{dt},\]
\[mult(x_{1})\rightarrow id,\;\frac{\partial}{\partial x_{1}}\rightarrow
\beta - t\frac{d}{dt}.\]
For $\gtsl_{2}$ there is essentially 1 Casimir operator
\[\Omega = EF+FE+\frac{1}{2}H^{2}\]
and one can easily write down the equation (\ref{diffeqclassgen}) in this case.
Before doing this we
remark that the following additional simplification appears here. The element
$\Omega$ may be
regarded as a $\gtsl_{2}-$invariant element of the symmetric square
$S^{2}\gtsl_{2}$.
The latter algebra acts on a tensor product of a pair of $\gtsl_{2}-$modules by
the formula simpler
than that for $U(\gtsl_{2})$:
\[X\otimes X\,v\otimes w=Xv\otimes Xw.\]
This gives another action of $\Omega$ and arguments which led to
(\ref{diffeqclassgen}) apply to
this action as well. Taking into account all this one finds that the function
$\widetilde{F^{\alpha}x_{1}^{\lambda_{1}}x_{2}^{\lambda_{2}}}(t)$
satisfies the classical hypergeometric equation
\begin{equation}
\label{diffeqclasssl2}
t(t-1)x(t) ''
+((1-\lambda_{1}-2\lambda_{2}+\alpha)t-\lambda_{1}-\lambda_{2}+\alpha-1)x(t)
'+\lambda_{2}(\lambda_{1}+\lambda_{2}-\alpha)x(t)=0,
\end{equation}
which is by no means surprise in view of sect. \ref{ExplicitFormulasforgtsl2}.

3. {\em  Example: $U_{q}(\gtsl_{2})$.} Of course all the arguments which led to
(\ref{diffeqclassgen}) apply to the case of a ``finite-dimensional'' quantum
group and given its
realization by $q-$difference operators one obtains a system of $q-$difference
equations on a certain
function. Here we are able to carry out the necessary calculations for
$U_{q}(\gtsl_{2})$. In this
case we set $\cm_{q}(\lambda)$ for all $f(x)$ to be a suitable family of
functions on $\nc$, action of $U_{q}(\gtsl_{2})$ being given by
(\ref{realsl2q_1} -
\ref{realsl2q_3}). Relations(\ref{realsl2q_1} -
\ref{realsl2q_3}) imply that  $x^{2\lambda}\in \cm_{q}(\lambda)$ is the highest
weight vector.
 The Casimir operator for $U_{q}(\gtsl_{2})$ is given by
\[\Omega = FE+\frac{Kq+K^{-1}q^{-1}}{(q-q^{-1})^{2}}.\]
As above one finds that the function
$F^{\alpha}\,x_{1}^{2\lambda_{1}}x_{2}^{2\lambda_{2}}$, which is
understood as an element of the module $\cm_{q}(\lambda_{1})\otimes
\cm_{q}(\lambda_{2})$,
 satisfies
\begin{equation}
\label{qdiffeqclasssl2}
\Omega f(x_{1},x_{2})=\frac{q^{2(\lambda_{1}+\lambda_{2})+1}+
q^{-2(\lambda_{1}-\lambda_{2})+1}}{(q-q^{-1})^{2}}f(x,y).
\end{equation}
Again using the fact that the function in question is homogeneous one makes it
into a function in 1
variable, the $q-$ difference operator in 2 variables representing $\Omega$
being transformed in an
obvious way into a $q-$difference operator in 1 variable. Finally what one
obtains is the usual
$q-$hypergeometric equation, which is again no surprise in view of the formula
 (\ref{eva_of_q_compow_m_F} ).

\end{document}